\documentclass[useAMS,usenatbib]{mn2e}
\usepackage{natbib}
\usepackage{deluxetable}
\usepackage{threeparttable}
\usepackage{longtable}
\usepackage{environ}
\usepackage{graphicx}
\usepackage{textcomp}
\usepackage[fleqn]{amsmath}
\usepackage{lscape}
\usepackage{pdflscape}
\usepackage{dcolumn}
\usepackage{rccol}
\usepackage{threeparttablex}
\usepackage{amssymb}
\usepackage{comment} 
\usepackage{xcolor}

\newcommand{\ugriz}{\protect\hbox{$ugriz$} }
\newcommand{\grizy}{\protect\hbox{$grizy$} }
\newcommand{\uvgriz}{\protect\hbox{$uvgriz$} }

\newcommand{\Siii}{\ensuremath{\mathrm{Si}\,\textsc{ii}\,\lambda6355}}
\newcommand{\Siiitmp}{\ensuremath{\mathrm{Si}\,\textsc{ii}\,\lambda5972}}

\newcommand{\deltam}{\ensuremath{\Delta m_{15}}}

\title[Swift SNe~Ia]{Swift UVOT Grism Observations of Nearby Type~Ia Supernovae -- II. Probing the Progenitor Metallicity of SNe~Ia with Ultraviolet Spectra}
\author[Pan et al.]{
Y.-C.~Pan$^{1,2}$\thanks{E-mail:yenchen.pan@nao.ac.jp},
R.~J.~Foley$^{3}$,
D.~O.~Jones$^{3}$,
A.~V.~Filippenko$^{4,5}$,
N.~P.~M.~Kuin$^{6}$
\\
  $^{1}$Division of Science, National Astronomical Observatory of Japan, 2-21-1 Osawa, Mitaka, Tokyo 181-8588, Japan\\
  $^{2}$EACOA Fellow\\
  $^{3}$Department of Astronomy and Astrophysics, University of California, Santa Cruz, CA 95064, USA\\
  $^{4}$Department of Astronomy, University of California, Berkeley, CA 94720-3411, USA\\
  $^{5}$Miller Senior Fellow, Miller Institute for Basic Research in Science, University of California, Berkeley, CA 94720, USA\\
  $^{6}$Mullard Space Science Laboratory/University College London, Holmbury St. Mary, Dorking, Surrey, RH5 6NT, UK\\
}

\begin{document}

\maketitle

\label{firstpage}

\begin{abstract}
Ultraviolet (UV) observations of Type Ia supernovae (SNe~Ia) are crucial for constraining the properties of their progenitor systems. Theoretical studies predicted that the UV spectra, which probe the outermost layers of a SN, should be sensitive to the metal content of the progenitor. Using the largest SN~Ia UV ($\lambda<2900$\,\AA) spectroscopic sample obtained from {\it Neil Gehrels Swift Observatory}, we investigate the dependence of UV spectra on metallicity. For the first time, our results reveal a correlation ($\sim2\sigma$) between SN~Ia UV flux and host-galaxy metallicities, with SNe in more metal-rich galaxies (which are likely to have higher progenitor metallicities) having lower UV flux level. We find that this metallicity effect is only significant at short wavelengths ($\lambda\lesssim2700$\,\AA), which agrees well with the theoretical predictions. We produce UV spectral templates for SNe~Ia at peak brightness. With our sample, we could disentangle the effect of light-curve shape and metallicity on the UV spectra. We also examine the correlation between  the UV spectra and SN luminosities as parameterised by Hubble residuals. However, we do not see a significant trend with Hubble residuals. This is probably due to the large uncertainties in SN distances, as the majority of our sample members are extremely nearby (redshift $z\lesssim0.01$). Future work with SNe discovered in the Hubble flow will be necessary to constrain a potential metallicity bias on SN~Ia cosmology.
\end{abstract}


\section{Introduction}
\label{sec:introduction}

Type~Ia supernovae (SNe~Ia; see \citealt{1997ARA&A..35..309F} for a review of SN classification) are remarkable cosmological standardisable candles that are consistently used to measure cosmological parameters \citep[e.g.,][]{1998AJ....116.1009R,1999ApJ...517..565P,2007ApJ...659...98R,2009ApJS..185...32K,2012ApJ...746...85S,2014ApJ...795...44R,2014A&A...568A..22B,2016ApJ...826...56R,2018ApJ...853..126R,2018ApJ...859..101S}. Observations of SNe~Ia provide a direct route to probe the nature of the dark energy that drives the accelerated expansion. As these studies become increasingly more precise, systematic uncertainties become a significant component of the error budget \citep[e.g.,][]{2011ApJS..192....1C}. Thus, one important consideration for their future use is how well we can understand the effect of underlying systematics and their evolution with redshift.

Theoretical studies suggested the progenitor metallicity could be one potential systematic that significantly alters SN~Ia luminosities. \citet{2003ApJ...590L..83T} showed that the amount of $^{56}$Ni (which powers the SN light curve) produced during the SN explosion will strongly depend on the progenitor metallicity. SNe~Ia with higher progenitor metallicities will generate more $^{14}$N during the hydrogen-burning stage and then convert to neutron-rich $^{22}$Ne during the helium-burning stage. This $^{22}$Ne will favour the production of stable and neutron-rich $^{58}$Ni instead of the radioactive $^{56}$Ni, resulting in fainter peak luminosity. The evolution of this metallicity effect is believed to have substantial bias on SN~Ia cosmology \citep[e.g.,][]{2006astro.ph..8324P}.

Technically, the direct measurement of SN progenitor properties is extremely difficult. Indirect methods like host-galaxy studies have been proven to be a profitable way to probe the astrophysical effects in SN progenitors \citep[e.g.,][]{2010ApJ...715..743K,2010ApJ...722..566L,2010MNRAS.406..782S,2011ApJ...743..172D,2013ApJ...770..108C,2014MNRAS.438.1391P}. Some studies found that SN~Ia luminosities correlate with the host-galaxy metallicities even after the empirical corrections for light-curve width and colour \citep[e.g.,][]{2011ApJ...743..172D,2013ApJ...770..108C,2014MNRAS.438.1391P}, in the sense that brighter SNe~Ia (after corrections) tend to reside in more metal-rich galaxies. However, the host metallicities measured by these studies are mostly on a galactic scale. The discrepancy from the metallicity of the progenitor can be expected to deviate from the galactic mean. 

Another promising alternative to constrain the progenitor metallicity is through ultraviolet (UV) observations of SNe~Ia. Theoretical models suggested that the UV spectra of SNe~Ia should be sensitive to the progenitor metallicity. SNe with higher progenitor metallicity will increase the iron-group elements (IGEs) in the outer layers of the SN and cause greater UV line blanketing \citep[e.g.,][]{1998ApJ...495..617H,2000ApJ...530..966L}. Thus, we should see a more suppressed UV flux for SNe having higher progenitor metallicities. This metallicity effect will only dramatically change the UV spectrum but not the optical spectrum, as the electron scattering opacity (which is not sensitive to the line-blanketing effect) is a more dominant factor than the line opacity in the optical \citep[e.g.,][]{2008ApJ...674...51E}.

There have been many efforts to obtain UV spectra of SNe~Ia \citep[e.g.,][]{1993ApJ...415..589K,2008ApJ...674...51E,2008ApJ...686..117F,2012MNRAS.426.2359M}. However, most of the spectra from these studies do not probe $\lambda<2700$\,\AA\ (which by theory is most sensitive to the progenitor metallicity). Several studies were able to obtain high signal-to-noise ratio (SNR) UV spectra that probe shorter wavelengths \citep[down to $\sim1800$\,\AA; e.g.,][]{2012ApJ...753L...5F,2014MNRAS.439.1959M,2014MNRAS.443.2887F,2015MNRAS.452.4307P,2016MNRAS.461.1308F}, but the relatively small sample size makes them unable to cover the diversity of SNe~Ia and fully interpret the metallicity effect.

Recent observations of ``twin'' SN~2011by and SN~2011fe provide some evidence of a potential metallicity effect on SN~Ia UV spectra \citep{2013ApJ...769L...1F,2015MNRAS.446.2073G,2018arXiv180608359F}. These two SNe have nearly identical optical light-curve widths and spectra but very different UV spectra. Intriguingly, despite being so similar in the optical, SN~2011fe and SN~2011by present significantly different peak absolute magnitudes \citep[$\Delta M_{V}=0.335\pm0.069$\,mag;][]{2018arXiv180608359F}. These studies suggested that the difference in UV flux between the two SNe is likely due to their significantly different progenitor metallicities. However, these twin SNe might be rare. They must have very similar explosions but with the variations in their UV spectra primarily driven by the metallicity difference. For two random SNe, we will need to disentangle the metallicity and other explosion effects when comparing their UV spectra. In theory, this can be done by generating a SN~Ia UV spectral template with which we can derive the relative metallicity of any SN~Ia without needing a twin, but a large UV spectral sample which covers a wide range of parameter space will be necessary to achieve this purpose.

Recently, \citet[][hereafter P18]{2018MNRAS.479..517P} constructed a large UV spectral sample with the {\it Neil Gehrels Swift Observatory} \citep{2004ApJ...611.1005G}. Their study contains 120 UV spectra of 39 SNe~Ia, representing the largest UV ($\lambda<2900$\,\AA) spectroscopic sample of SNe~Ia to date. However, owing to the slitless design of the {\it Swift} UV grism, the target spectrum is more easily contaminated by nearby background sources. P18 developed an improved data-reduction procedure to produce more-accurate spectra. With this improved sample, we can probe a larger parameter space of SNe~Ia and construct better UV spectral templates than previous studies. In this second paper of the series, we perform detailed analysis of the {\it Swift} UV spectroscopic sample with the goal of constraining the metallicity effect on SN~Ia UV spectra.

An outline of the paper follows. In Section~\ref{sec:data} we describe our SN~Ia sample and the host-galaxy determination. We present the data analysis in Section~\ref{sec:analysis}. Our results are discussed in Section~\ref{sec:discussion} and our conclusions are in Section~\ref{sec:conclusion}.

\section{Data}
\label{sec:data}
\subsection{SN sample}
\label{sec:sample}
\begin{figure}
	\centering
		\includegraphics[scale=0.51]{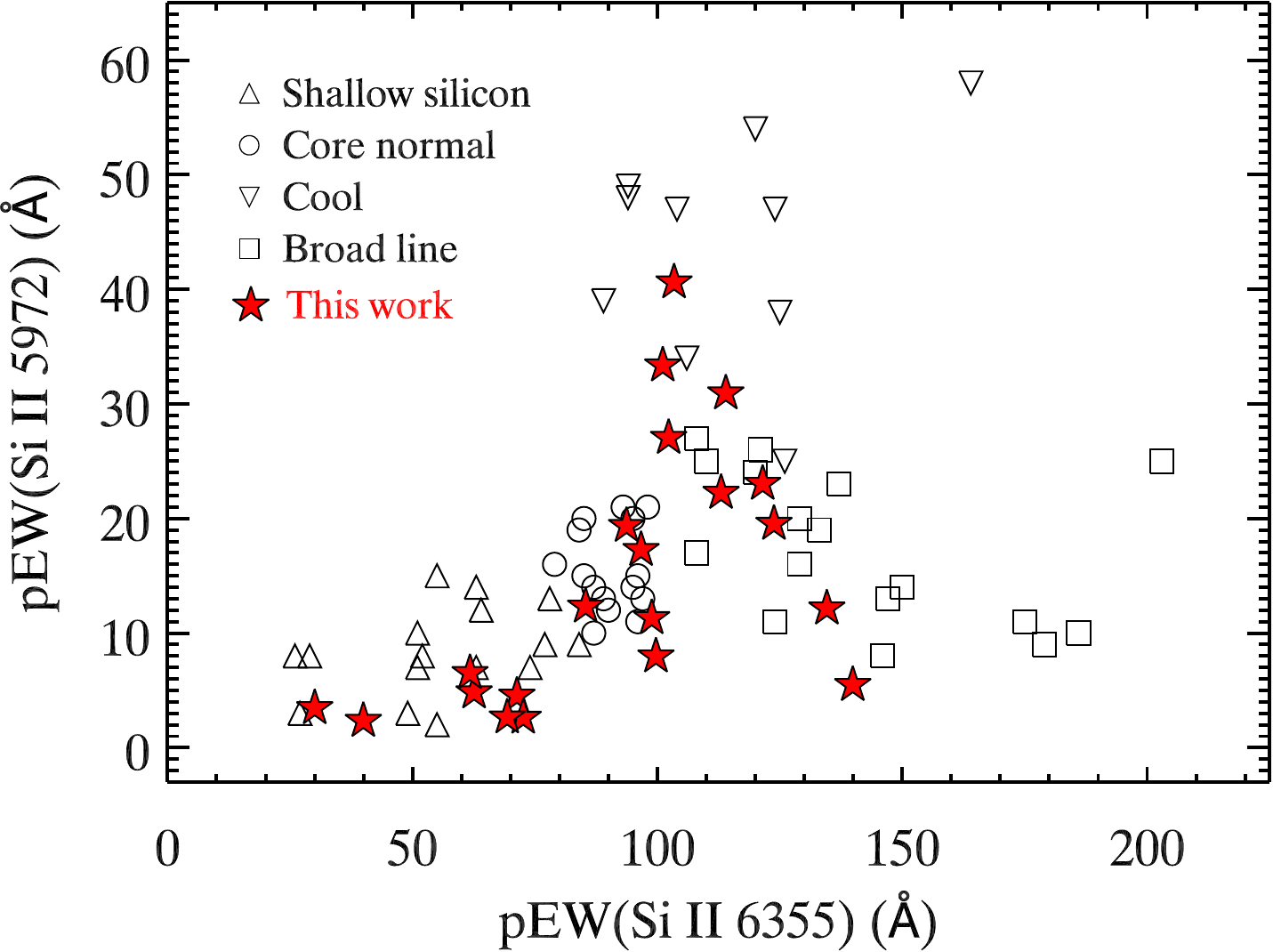}
		\caption{\Siiitmp\ pEW as the function of \Siii\ pEW.
               The sample and subclasses studied by \citet{2009PASP..121..238B}
               is plotted. The sample is split into four
               different subclasses: ``shallow silicon'' (open triangles),
               ``core normal'' (open circles), ``cool'' (open downward triangles),
               and ``broad line'' (open squares). The position of the near-peak sample
               in this work is overplotted (represented by the red filled stars).}
        \label{spec_subclass}
\end{figure}

\begin{figure}
	\centering
		\includegraphics[scale=0.46]{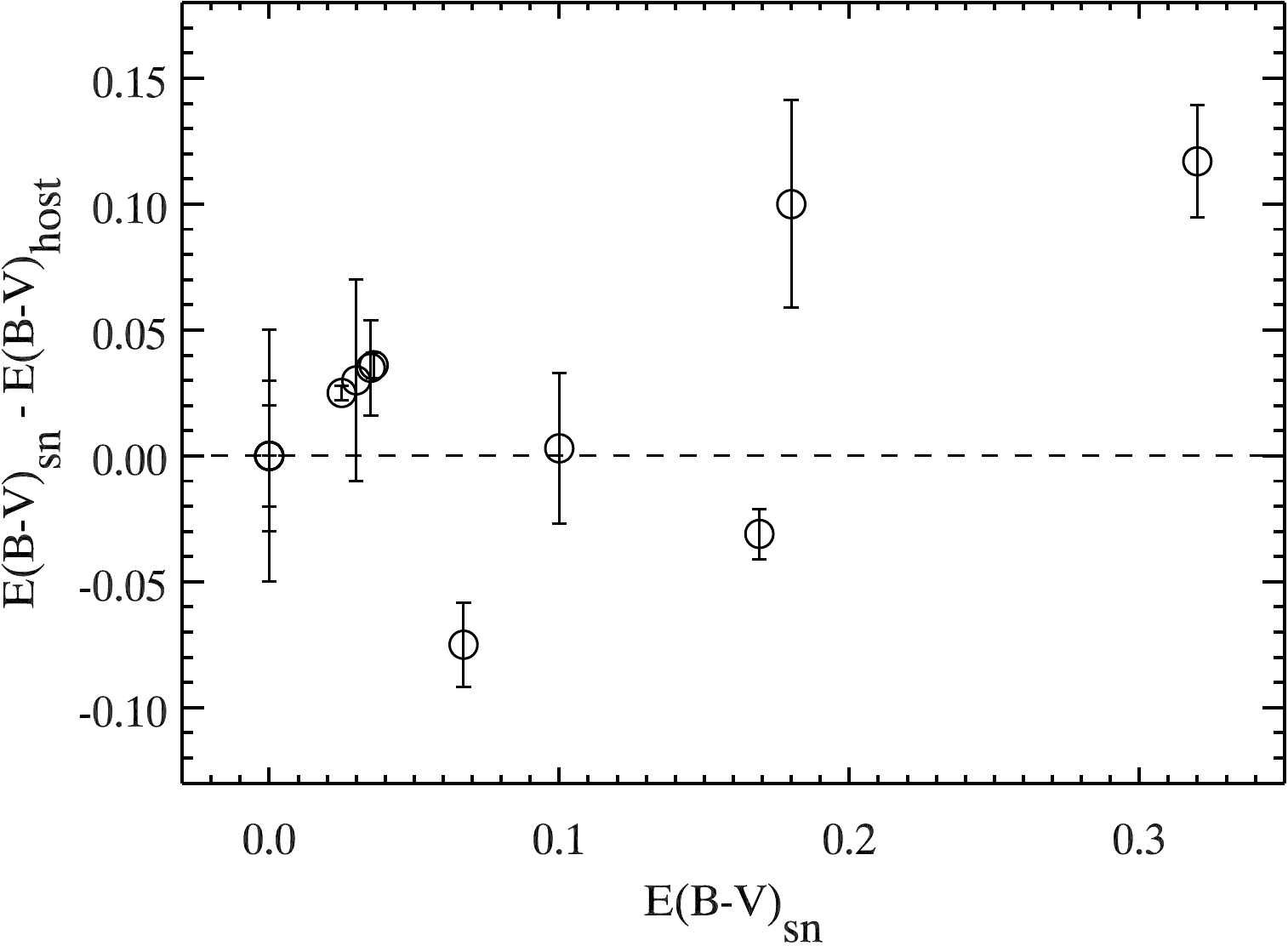}
		\caption{The comparison between the colour excess derived from the host-galaxy spectra ($E(B-V)_{\rm host}$) and SN light-curves ($E(B-V)_{\rm SN}$). Here we plot the difference between $E(B-V)_{\rm SN}$ and $E(B-V)_{\rm host}$ as the function of $E(B-V)_{\rm SN}$.}
        \label{ebv_compare}
\end{figure}

\begin{table*}
\caption{Summary of the SN~Ia near-peak sample in this work.}
\begin{tabular}{lccrcccrrc}
\hline\hline
SN Name         & Redshift $z$ & \deltam($B$)$^a$ & Phase & Host name & $E(B-V)_{\rm host}$ & Morph.$^b$ & log\,$M_{\rm stellar}^c$ & 12 + log\,(O/H)$^d$ & AGN?$^e$\\
                &          &   (mag)          & (day) &      &       (mag)         &        &     (M$_{\odot}$)     &                   &          \\
\hline
SN~1992A$^f$    & 0.0062 & 1.47(05) & $4.9$  & NGC~1380      & 0.00(02)                & S0  & $11.451^{+0.033}_{-0.046}$ & $8.811^{+0.005}_{-0.005}$ & Y\\
SN~2005cf       & 0.0065 & 1.05(03) & $-1.0$ & MCG-01-39-03  & 0.10(03)                & S0  & $9.357^{+0.059}_{-0.200}$  & \nodata                   & N\\
SN~2005df       & 0.0044 & 1.12(00) & $0.7$  & NGC~1559      & 0.00(02)                & Sc  & $10.693^{+0.066}_{-0.827}$ & \nodata                   & \nodata\\
SN~2005ke       & 0.0049 & 1.76(01) & $-2.7$ & NGC~1371      & 0.04(01)                & Sa  & $10.283^{+0.495}_{-0.161}$ & \nodata                   & N\\
SN~2008Q        & 0.0079 & 1.41(05) & $-0.2$ & NGC~524       & 0.00(02)                & S0  & $10.572^{+0.380}_{-0.346}$ & \nodata                   & \nodata\\
SN~2009an       & 0.0092 & 1.44(00) & $2.7$  & NGC~4332      & 0.09(03)                & Sa  & $11.001^{+0.001}$          & $8.872^{+0.018}_{-0.024}$ & N\\
SN~2009ig       & 0.0088 & 0.89(00) & $1.3$  & NGC~1015      & 0.00(05)                & Sa  & $10.435_{-0.175}$          & \nodata                   & N\\
SN~2011ao       & 0.0107 & 1.00(00) & $-2.0$ & IC~2973       & 0.04(02)                & Scd & $9.854^{+0.117}_{-0.033}$  & $8.774^{+0.001}_{-0.000}$ & N\\
SN~2011by       & 0.0028 & 1.14(03) & $0.6$  & NGC~3972      & 0.07(01)                & Sb  & $9.294^{+0.028}_{-0.191}$  & $8.795^{+0.151}_{-0.087}$ & N\\
SN~2011fe       & 0.0008 & 1.11(00) & $0.6$  & NGC~5457      & 0.03(00)                & Sc  & \nodata                    & $8.717^{+0.003}_{-0.001}$ & N\\ 
SN~2011iv       & 0.0065 & 1.69(05) & $0.5$  & NGC~1404      & 0.00(03)                & E   & $11.407^{+0.024}_{-0.603}$ & \nodata                   & N\\
SN~2012cg$^f$   & 0.0014 & 0.86(00) & $-0.5$ & NGC~4424      & 0.18(04)                & Sa  & $8.793^{+0.063}_{-0.132}$  & $8.820^{+0.023}_{-0.009}$ & N\\
SN~2012fr       & 0.0055 & 0.85(05) & $0.9$  & NGC~1365      & 0.03(04)                & Sb  & $10.363^{+1.273}$          & $8.604^{+0.000}_{-0.001}$ & N\\
SN~2012ht       & 0.0036 & 1.39(05) & $-3.1$ & NGC~3447      & 0.00(02)                & Sm  & $6.516^{+0.278}_{-0.376}$  & $8.589^{+0.102}_{-0.112}$ & N\\
SN~2013aa       & 0.0040 & 0.80(03) & $-1.3$ & NGC~5643      & 0.00(05)$^g$            & Sc  & $10.474^{+0.086}_{-0.129}$ & $8.465^{+0.001}_{-0.001}$ & Y\\
SN~2013dy$^f$   & 0.0039 & 0.92(01) & $-0.8$ & NGC~7250      & 0.21(01)                & Scd & $8.752^{+0.034}_{-0.180}$  & \nodata                   & N\\
iPTF14bdn       & 0.0156 & 0.84(05) & $5.0$  & UGC~8503      & 0.00(05)$^g$            & Irr & $9.223^{+0.046}_{-0.124}$  & $8.437^{+0.023}_{-0.025}$ & N\\
ASASSN-14lp$^f$ & 0.0051 & 0.80(01) & $0.1$  & NGC~4666      & 0.32(01)                & Sc  & $11.072^{+0.000}_{-0.000}$ & $8.682^{+0.033}_{-0.037}$ & N\\
SN~2015F$^f$    & 0.0049 & 1.26(10) & $-0.3$ & NGC~2442      & 0.04(03)                & Sbc & $10.559^{+0.082}_{-0.163}$ & \nodata                   & \nodata\\
SN~2016ccz      & 0.0150 & 1.00(02) & $-1.2$ & MRK~685       & 0.00(05)$^g$            & Sb  & $9.044^{+0.022}_{-0.059}$  & $8.295^{+0.030}_{-0.036}$ & N\\
SN~2016coj      & 0.0045 & 1.33(03) & $-2.1$ & NGC~4125      & 0.03(05)$^g$            & E   & $10.794^{+0.406}_{-0.048}$ & \nodata                   & Y\\
SN~2016eiy      & 0.0087 & \nodata  & $0.6$  & ESO~509-IG064 & 0.00(05)$^g$            & Sab & $8.716^{+0.718}_{-0.258}$  & $8.540^{+0.001}_{-0.001}$ & N\\
SN~2016ekg      & 0.0171 & 1.00(00)  & $-0.7$ & PGC~67803     & 0.00(05)$^g$            & Sbc & $8.466^{+0.905}_{-0.881}$  & $8.490^{+0.001}_{-0.000}$ & N\\
SN~2016fff      & 0.0114 & 1.49(05) & $2.4$  & UGCA~430      & 0.00(02)                & Sd  & $8.267^{+0.954}_{-0.735}$  & $8.501^{+0.002}_{-0.001}$ & N\\
SN~2017cbv      & 0.0040 & 1.06(00) & $-4.6$ & NGC~5643      & 0.00(02)                & Sc  & $10.474^{+0.086}_{-0.129}$ & \nodata                   & \nodata\\
SN~2018aoz      & 0.0058 & 1.12(00) & $-1.2$  & NGC~3923     & 0.00(02)                & E   & $8.857^{+1.822}_{-0.118}$  & \nodata                   & N\\
\hline
\label{peak-sample}
\end{tabular}
\begin{flushleft}
$^a${The $B$-band decline 15~days after the peak brightness.}\\
$^b${Host-galaxy morphology from HyperLeda \citep{2014A&A...570A..13M}.}\\
$^c${Host-galaxy stellar mass.}\\
$^d${Host-galaxy gas-phase metallicity based on the PP04 O3N2 calibration \citep{2004MNRAS.348L..59P}.}\\
$^e${Is the host galaxy contaminated by an AGN?}\\
$^f${UV spectra observed by {\it HST}/STIS.}\\
$^g${$E(B-V)$ determined by fitting the host-galaxy spectrum.}\\
\end{flushleft}
\end{table*}

\begin{figure*}
	\centering
		\includegraphics[scale=0.49]{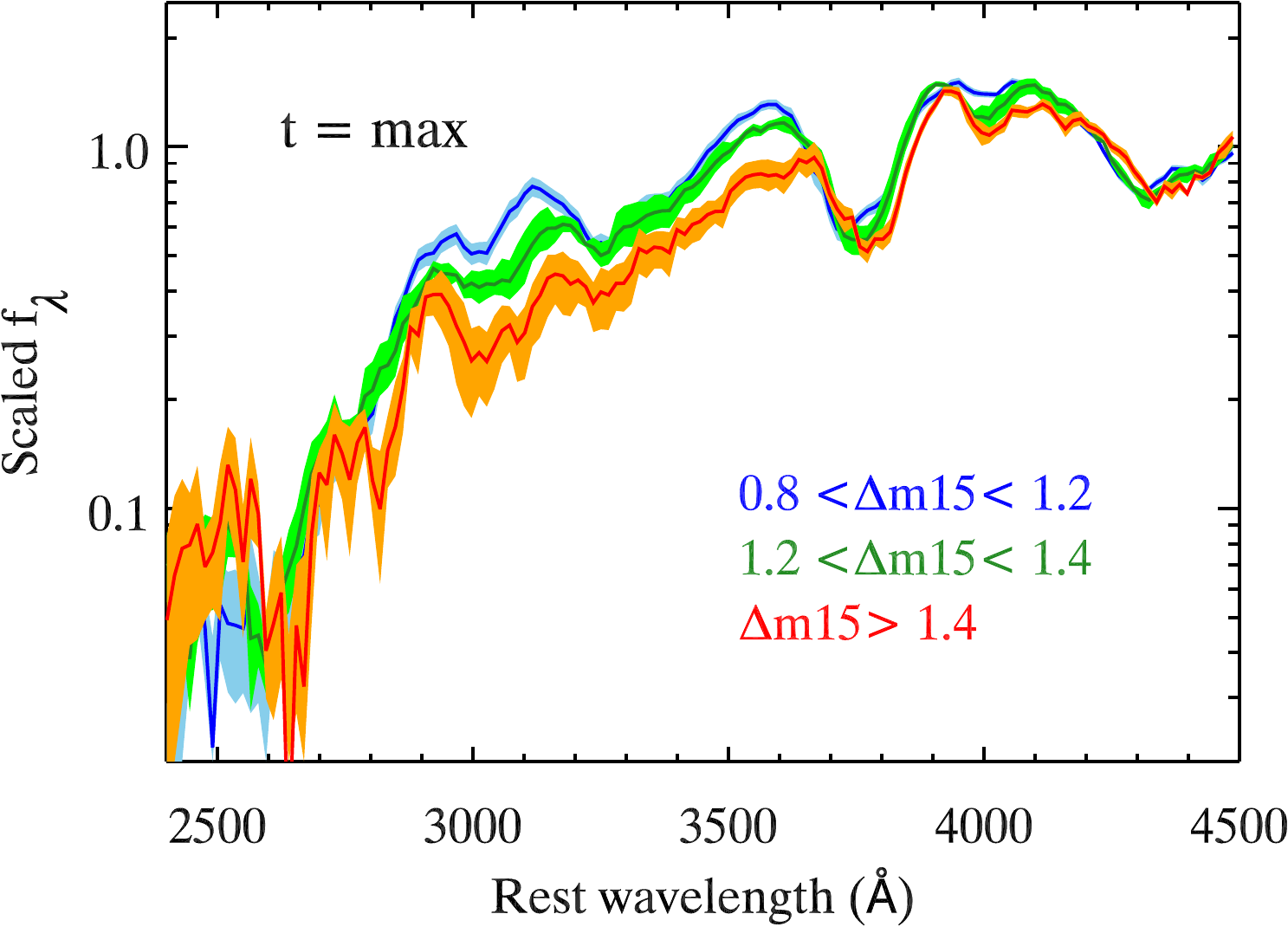}
		\hspace{0.25cm}
		\includegraphics[scale=0.49]{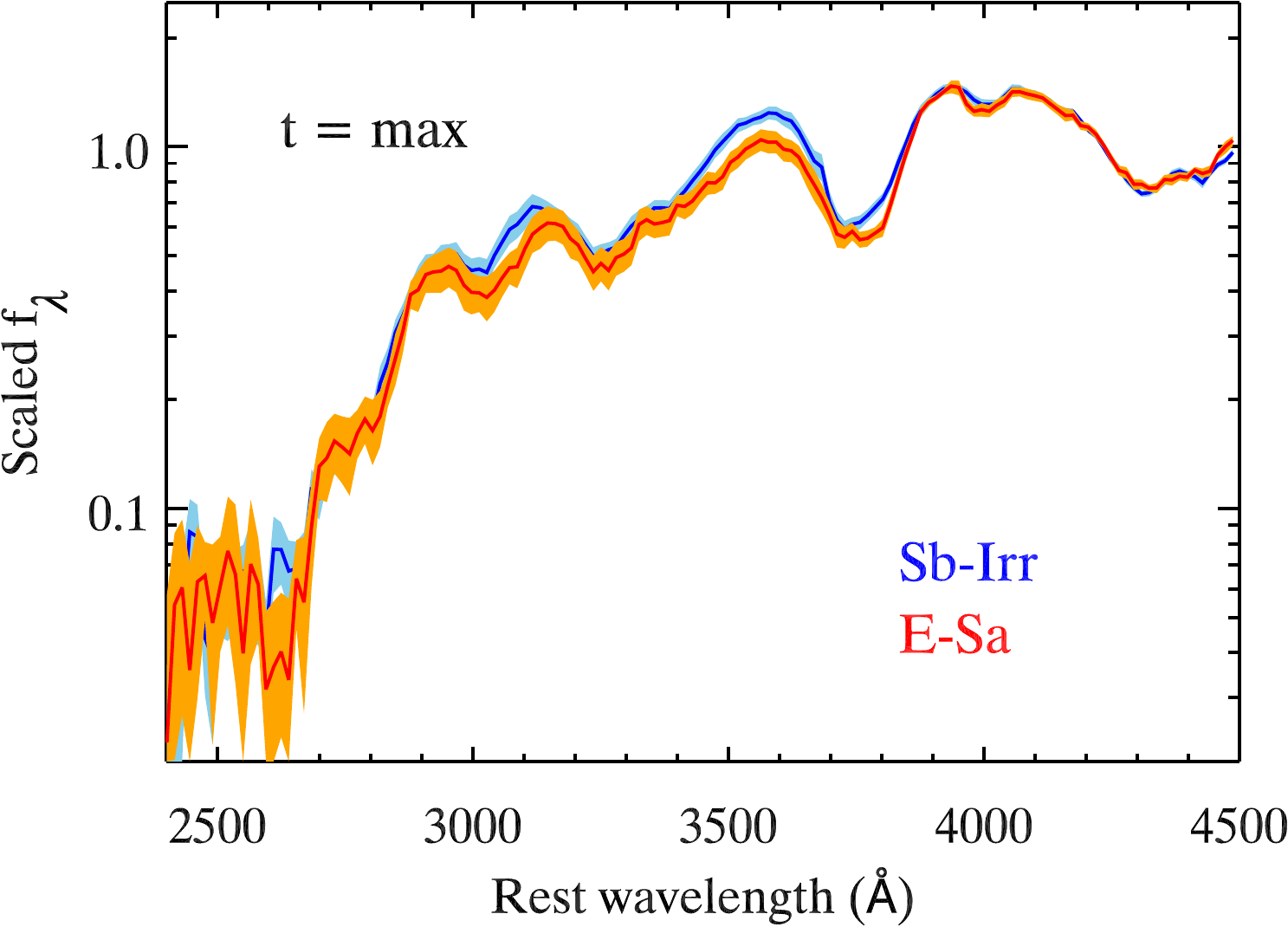}\\
		\vspace{0.25cm}
		\includegraphics[scale=0.49]{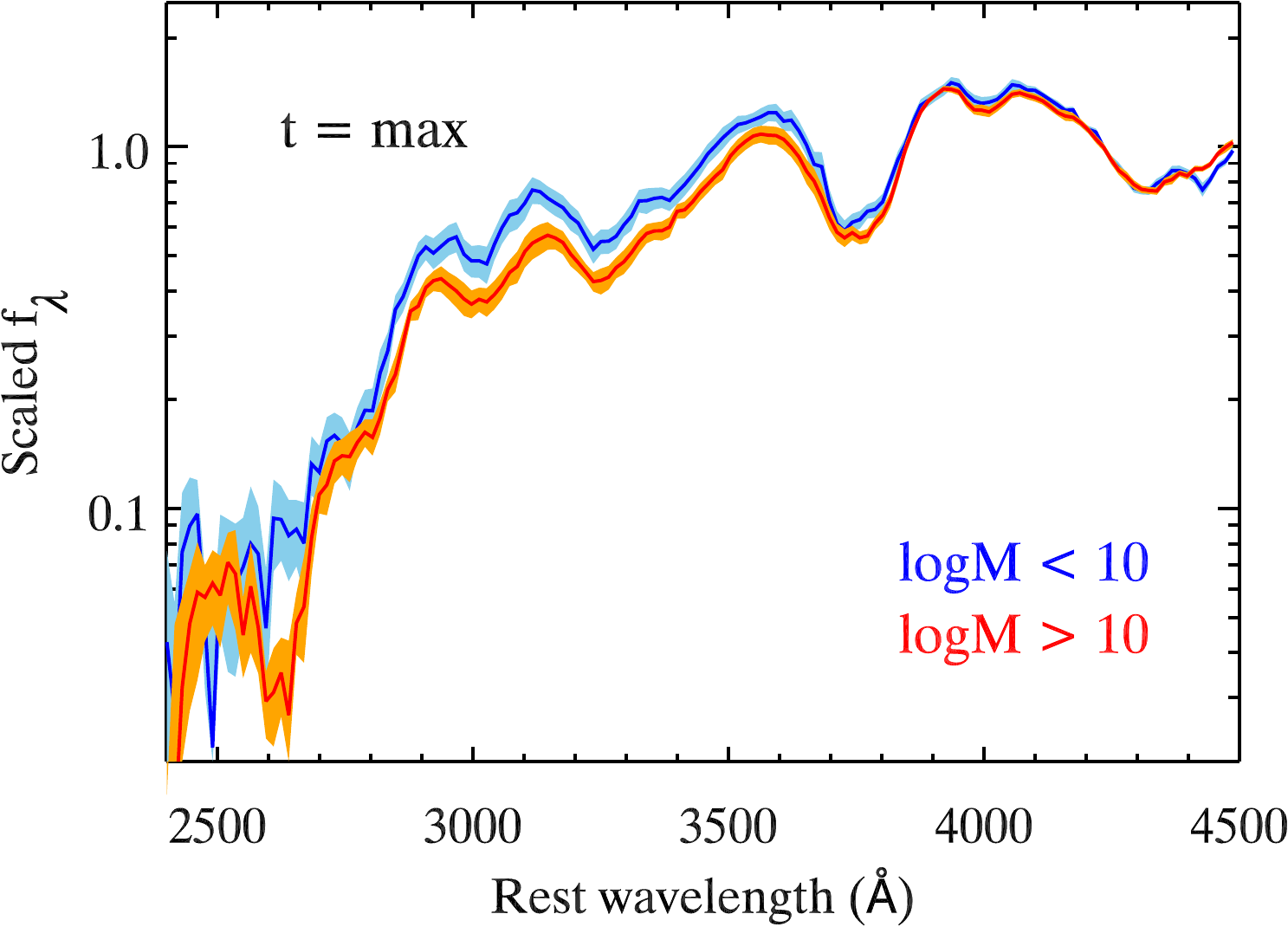}
		\hspace{0.25cm}
		\includegraphics[scale=0.49]{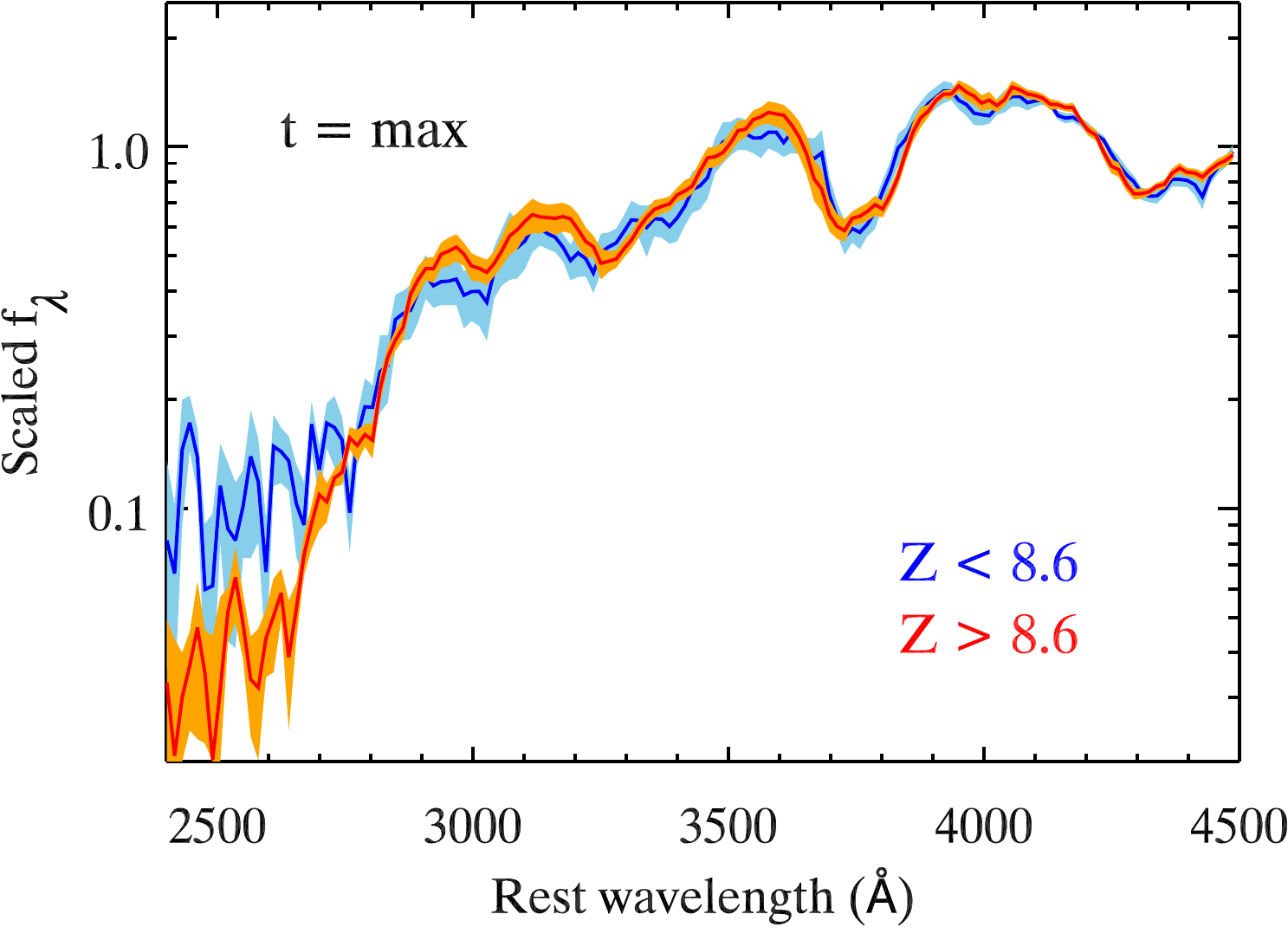}
                \caption{The mean spectrum of the {\it Swift} near-peak sample with respect to $\deltam$($B$) of SNe~Ia (upper left), host morphology (upper right), host stellar mass (lower left), and host gas-phase metallicity (lower right), respectively. The spectra are colour-coded based on the parameter range. The shaded regions represent the 68.27\% confidence region from the bootstrap resampling. All of the spectra are rebinned with a bin size of 15\,\AA.}
        \label{mean-spec}
\end{figure*}

A detailed description of the {\it Swift} data can be found in P18 and has been summarised below, along with any of the additions made to define our sample.

Our original sample contains 120 {\it Swift}/UVOT grism spectra of 39 SNe Ia. The redshift of that sample ranges from $z = 0.0006$ to 0.0214 (a median of 0.0079). The $\deltam$($B$) (defined as $B$-band decline 15~days after peak $B$-band brightness) of that sample ranges from 0.6\,mag to 1.8\,mag, with three more SNe measured ($\deltam(B) = 1.00$, 1.11, and 1.12\,mag for SN~2016ekg, SN~2017erp, and SN~2018aoz, respectively) since P18\footnote{The $\deltam(B)$ of SN~2016ekg and SN~2017erp are determined from {\it Swift} UVOT $B$-band light curves. The $\deltam(B)$ of SN~2018aoz is determined from a Swope $B$-band light curve.}.

Two-thirds of the SNe in our sample have multiple epochs of UV spectra, with a median phase of $-4.5$ and $-1.9$\,days for the first observation of a SN and all spectra, respectively. In this work, we focus our major analysis on the near-peak-brightness spectra (i.e., within 5\,days relative to the maximum light). This contains 28 SNe from our parent sample in P18. 

Among the 28 SNe, we exclude SN~2014J owing to its high and uncertain reddening \citep[e.g.,][]{2014ApJ...784L..12G,2014MNRAS.443.2887F,2014ApJ...788L..21A}. The super-Chandra SN~2012dn is also excluded owing to its anomalous behaviour in the UV \citep{2014ApJ...787...29B}. Five SNe (SN~2008hv, SN~2009Y, SN~2010ev, SN~2013cg, and SN~2016gsb) were further excluded owing to either poor data quality or insufficient wavelength coverage for our analysis. To increase the sample size, we include 5 SNe (SN~1992A, SN~2012cg, SN~2013dy, ASASSN-14lp, and SN 2015F) which do not have near-peak {\it Swift}/UVOT grism spectra but rather near-peak {\it HST}/STIS UV spectra. We also include all other {\it HST} UV spectra studied by \citet{2016MNRAS.461.1308F} to our analysis when appropriate. Our final near-peak sample contains 26 SNe (see Table~\ref{peak-sample}).

Figure~\ref{spec_subclass} shows the subclasses of SNe~Ia according to the pseudo-equivalent width (pEW) of \Siii\ and \Siiitmp\ \citep{2006PASP..118..560B}. \citet{2006PASP..118..560B} split the SN~Ia sample into four groups: A ``shallow-silicon'' group, which have low pEWs for both \Siii\ and \Siiitmp\ lines (and include SN~1991T-like objects); a ``core-normal'' group, which have homogeneous and intermediate pEWs; a ``cool'' group \citep[similar to the FAINT group in][]{2005ApJ...623.1011B}, with strong \Siiitmp\ lines relative to \Siii; and a ``broad-line'' group, which present strong \Siii\ absorption lines. We also display our near-peak sample (where we were able to measure the pEWs from the optical spectra) in Figure~\ref{spec_subclass}.  It is clear that our sample covers all four subclasses and probes a larger parameter space (especially those in the broad-line group) than that of the {\it HST} UV sample of \citet{2016MNRAS.461.1308F}.

\subsection{UV Spectra}
\label{sec:spectra}
The spectroscopic observations were performed by the {\it Swift} UVOT \citep{2004SPIE.5165..262R,2015MNRAS.449.2514K}. Owing to the slitless design of the {\it Swift} UV grism observations, the target spectrum is easily contaminated by nearby background sources (e.g., host galaxy). Thus, we have developed an improved data-reduction procedure to mitigate this issue. A complete description of our methods can be found in P18. The details of the $\it HST$ UV spectra included in our sample are presented by \citet{2016MNRAS.461.1308F}.

We correct all spectra for foreground Galactic reddening using the calibrations of \citet{2011ApJ...737..103S}. We use the host-galaxy reddening determined from the SN light curves as reported by previous studies \citep{1999AJ....118.1766P,2009ApJ...697..380W,2010AJ....139..120F,2012ApJ...744...38F,2012ApJ...753L...5F,2013NewA...20...30M,2013MNRAS.430..869S,2014ApJ...782L..35Y,2015MNRAS.452.4307P,2015ApJS..221...22I,2016ApJ...826..144S,2017ApJ...845L..11H,2016ApJ...820...92M,2017RNAAS...1a..36K,2018ApJ...864L..35S,2018PASP..130k4504P}. Here we adopt $R_{V}=3.1$ and a \citet*[][CCM]{1989ApJ...345..245C} reddening law. Some of these SNe (4 out of 18) do not have uncertainties measured for host-galaxy reddening. Here we adopt a typical error of 0.02\,mag in their host $E(B-V)$, as determined from the mean value of those having uncertainties measured from their light curves in our sample. However, we find that our result is not sensitive to the assigned uncertainty or the choice of reddening laws (see Section~\ref{sec:flux-ratio} for a discussion).

For SNe without host extinction derived from their light curves (only 6 SNe in our near-peak sample), we dereddened their spectra with the colour excess given by the host spectrum fitting. This is done with the reddening model in \textsc{gandalf} (see Section~\ref{sec:host} for more details). By comparing the observed spectra to the unreddened spectral templates, \textsc{gandalf} can determine the diffusive dust throughout the whole galaxy that affects the entire host spectrum including emission lines and the stellar continuum. We compare the colour excess $E(B-V)$ derived from host-galaxy spectral fitting and SN light-curve fitting for those SNe having both measurements available; the result is shown in Figure~\ref{ebv_compare}. We find that the $E(B-V)$ values determined from the two different methods generally agree well with each other, with a dispersion of only $\lesssim0.05$\,mag. The dispersion becomes even smaller for those objects suffering lower extinction. This is likely to be the case for the 6 SNe measured with host-galaxy spectral fitting in our near-peak sample, where we find no or negligible extinctions from their host galaxies. We have incorporated this dispersion in estimating the uncertainty of host-galaxy reddening (for those determined from the host spectra).

Two SNe (SN~2016fff and SN~2018aoz) in our near-peak sample do not have host reddening derived from either method; we assume no host-galaxy extinction given their position in the outskirts of their host galaxies. The $E(B-V)$ of each host galaxy in the near-peak sample is listed in Table~\ref{peak-sample}.

\subsection{Host-galaxy properties}
\label{sec:host}
To investigate the relation between SN~Ia UV spectrum and host galaxy, we determine the host-galaxy parameters (e.g., morphology, stellar mass, and metallicity) with both photometric and spectroscopic data. 

The morphology of the host galaxy is given by HyperLeda\footnote{http://leda.univ-lyon1.fr/} \citep{2014A&A...570A..13M}, with the numerical code for the host morphology defined by de Vaucouleurs morphological type in RC2 \citep{1976RC2...C......0D}. The host stellar mass is derived by fitting the multicolour photometry of the host galaxy with the photometric redshift code \textsc{z-peg} \citep{2002A&A...386..446L}. The host photometry is provided from SDSS \ugriz \citep{2018ApJS..235...42A}, Pan-STARRS1 \grizy \citep{2016arXiv161205560C}, and Skymapper \uvgriz \citep{2018PASA...35...10W} catalogs. \textsc{z-peg} fits the observed galaxy colours with galaxy spectral energy distribution (SED) templates corresponding to 9 spectral types (SB, Im, Sd, Sc, Sbc, Sb, Sa, S0, and E). We assume a \citet{1955ApJ...121..161S} initial-mass function (IMF).  A foreground dust screen varying from a colour excess of $E(B-V)=0$ to 0.2\,mag in steps of 0.02\,mag is used. The photometry is corrected for foreground Milky Way reddening with $R_{V} = 3.1$ and a CCM reddening law.

To measure the host metallicity, we obtained optical spectra of the host galaxies, primarily with the Kast spectrograph \citep{Kast_spectrograph} on the Lick Observatory 3\,m Shane telescope and the Goodman spectrograph \citep{2004SPIE.5492..331C} on the SOAR telescope. We fit the emission lines and stellar continuum of the host spectrum using the Interactive Data Language (\textsc{IDL}) codes \textsc{ppxf} \citep{2004PASP..116..138C} and \textsc{gandalf} \citep{2006MNRAS.366.1151S}. A complete description of this process can be found in \citet{2014MNRAS.438.1391P}. Briefly, \textsc{ppxf} fits the line-of-sight velocity distribution (LOSVD) of the stars in the galaxy in pixel space using a series of stellar templates. Before fitting the stellar continuum, the wavelengths of potential emission lines are masked to remove any possible contamination.  The stellar templates are based on the \textsc{miles} empirical stellar library \citep{2006MNRAS.371..703S, 2010MNRAS.404.1639V}.  A total of 288 templates is selected with $[M/H]=-1.71$ to $+0.22$ in 6 bins and ages ranging from $0.063$ to $14.12$\,Gyr in 48 bins. 

After the emission-line measurements from \textsc{ppxf} and \textsc{gandalf}, we calculate the host gas-phase metallicity with \textsc{pymcz} \citep{2016A&C....16...54B}, a python-based code that determines the oxygen abundances and their uncertainties with a Monte Carlo method. Here the PP04 O3N2 calibration \citep{2004MNRAS.348L..59P} is adopted. We further use BPT diagrams \citep*{1981PASP...93....5B} to check for potential contamination from active galactic nuclei (AGNs) in our host galaxies. The criteria proposed by \citet{2001ApJ...556..121K} are adopted to distinguish between normal and AGN host galaxies. For the host galaxies identified as AGNs, their gas-phase metallicity will be discarded from further analysis. The host-galaxy parameters measured for our near-peak sample can be found in Table~\ref{peak-sample}.

\begin{figure*}
	\centering
		\includegraphics[scale=0.74]{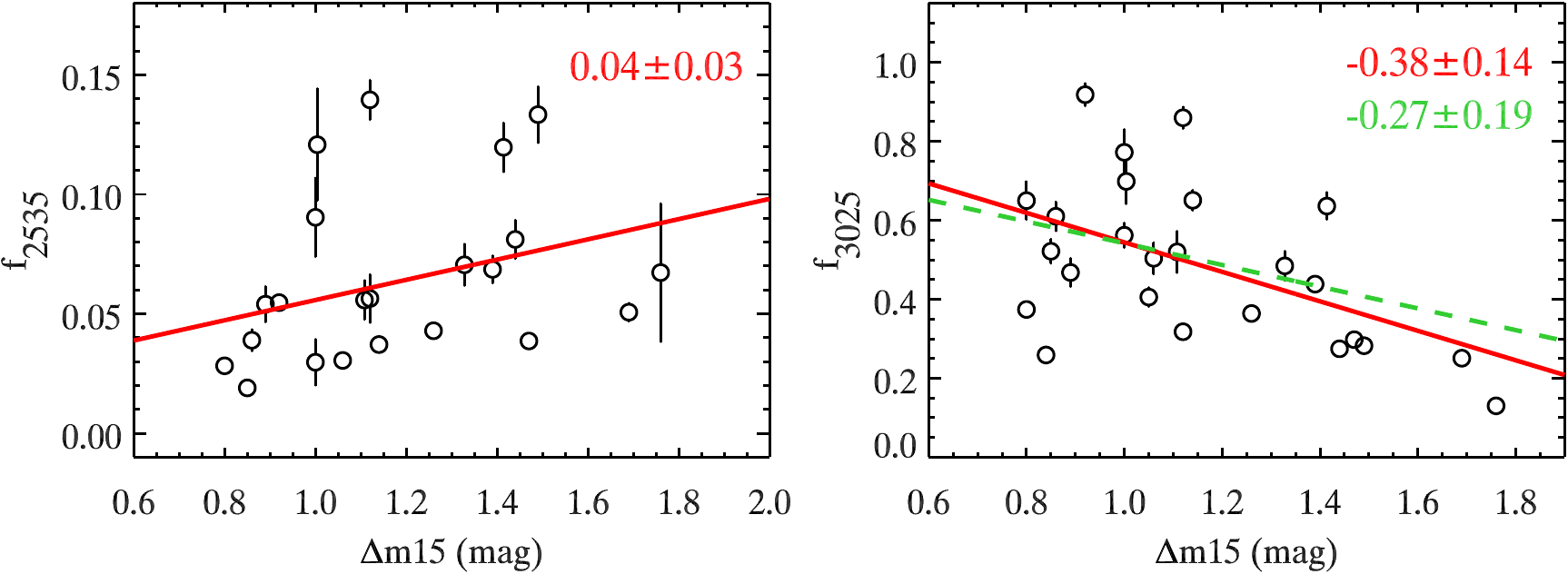}\\
		\vspace{0.1cm}
		\includegraphics[scale=0.74]{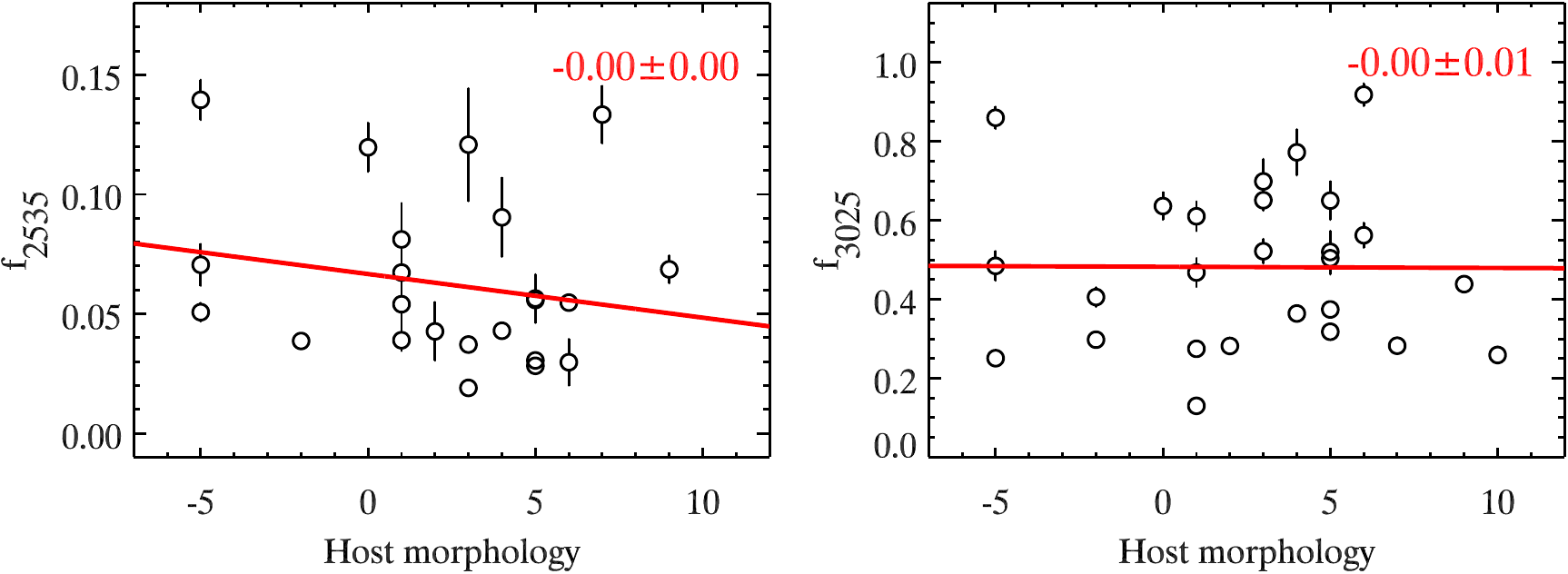}\\
		\vspace{0.1cm}
		\includegraphics[scale=0.74]{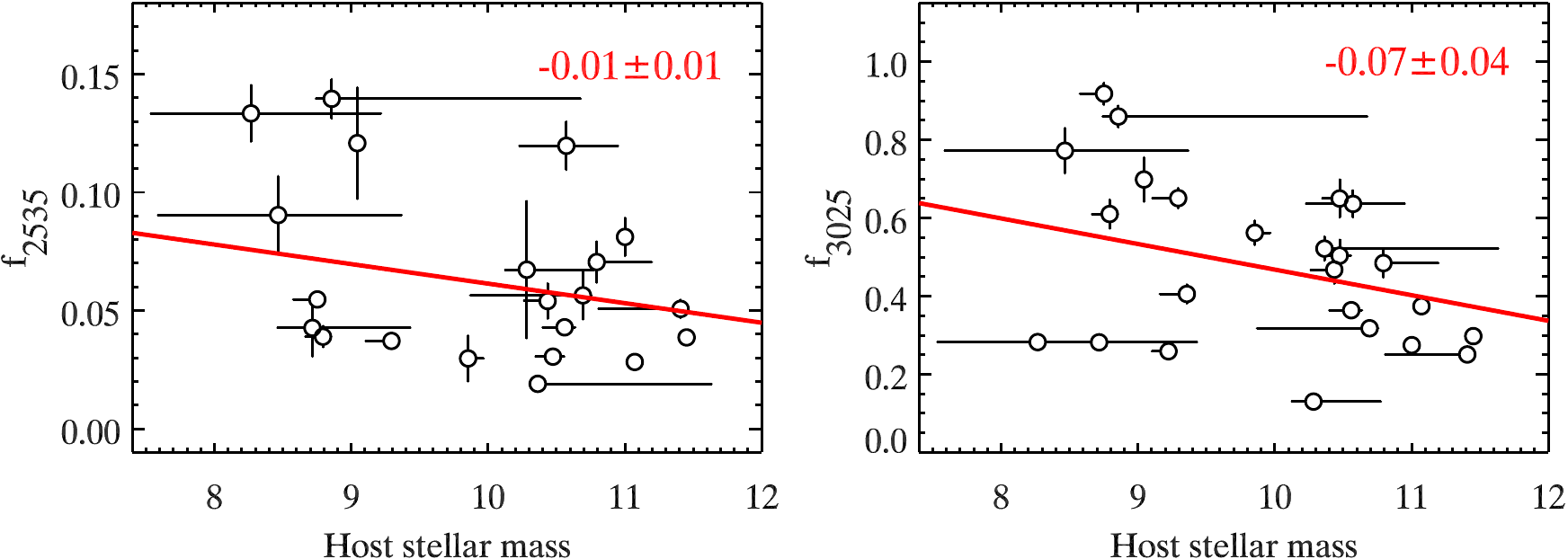}\\
		\vspace{0.1cm}
		\includegraphics[scale=0.74]{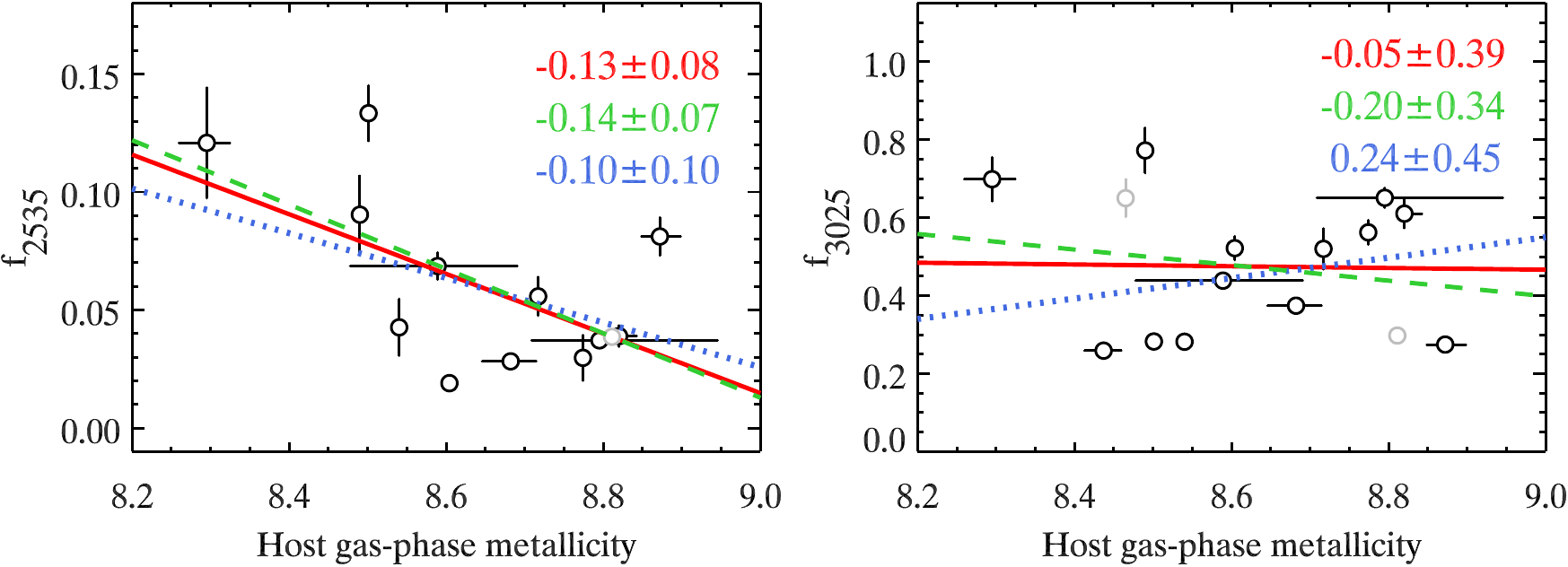}
                \caption{Flux ratios for 2450--2620 ($f_{2535}$, left panels) and 2900--3150\,\AA\ ($f_{3025}$; right panels) as a function of $\deltam$($B$) (top row), host morphology (second row), host stellar mass (third row), and host gas-phase metallicity (bottom row). The numerical code for the host morphology is defined by de Vaucouleurs morphological type in RC2 \citep{1976RC2...C......0D}. The solid line in each panel represents the best linear fit to the data. The dashed line in the top-right panel represents the linear fit with the two fastest decliners ($\deltam(B)>1.6$\,mag) excluded. The dashed lines and dotted lines in the bottom panels show the same linear fit, but including SNe in the potential AGN hosts (represented by grey circles) and excluding one SN having the most metal-poor host galaxy in our sample ($Z=8.267$), respectively. The legend in each panel indicates the slope of the fitted line.}
        \label{flux-ratio}
\end{figure*}

\begin{table*}
\centering
\caption{The trends of UV flux ratios with SN and host parameters.}
\begin{tabular}{lcccccccc}
\hline\hline
   & \multicolumn{4}{c}{$f_{2535}$} & \multicolumn{4}{c}{$f_{3025}$}\\
\hline
   & linear trend & probability of & \multicolumn{2}{c}{correlation} & linear trend & probability of & \multicolumn{2}{c}{correlation} \\
   &              & negative slope & Pearson & Kendall &              & negative slope & Pearson & Kendall \\ 
\hline
$\deltam$($B$) & $+0.043 \pm 0.030$ & 7.69\% & $+0.30$ & $+0.34$ & $-0.375 \pm 0.143$ & 99.4\% & $-0.52$ & $-0.35$\\
Morphology   & $-0.002 \pm 0.002$ & 79.1\% & $-0.18$ & $-0.12$ & $-0.002 \pm 0.011$ & 55.9\% & $-0.01$ & $+0.03$\\
Stellar Mass & $-0.009 \pm 0.007$ & 88.5\% & $-0.34$ & $-0.20$ & $-0.065 \pm 0.043$ & 94.0\% & $-0.33$ & $-0.30$\\
$Z$            & $-0.126 \pm 0.078$ & 94.9\% & $-0.60$ & $-0.27$ & $-0.051 \pm 0.389$ & 55.6\% & $-0.04$ & $+0.10$\\
$Z$ (incl. AGNs)& $-0.136 \pm 0.065$ & 97.9\% & $-0.66$ & $-0.32$ & $-0.204 \pm 0.336$ & 72.9\% & $-0.18$ & $-0.05$\\
\hline
\end{tabular}
\label{uvflux-statistic}
\end{table*}

\section{Analysis}
\label{sec:analysis}
\subsection{Mean spectra}
\label{sec:mean-spectra}
Here we produce the mean spectra to investigate potential correlation between the UV spectrum and other physical parameters, such as SN decline rate and host morphology, stellar mass, and gas-phase metallicity. The mean spectra are created using a bootstrap sampling method. The variance of the mean spectra is estimated with the range of spectra for the middle 68.27\% of spectra generated from the sample. All of the spectra are normalised with the median flux for a wavelength range of 4000--4500\,\AA. The results are shown in Figure~\ref{mean-spec}. 

We first examine the dependence of UV spectrum on $\deltam$($B$). Using {\it HST} spectra, \citet{2016MNRAS.461.1308F} found a strong trend between $\deltam$($B$) and the flux level at $\sim3000$--3500\,\AA, with higher $\deltam$($B$) (i.e., fast decliners) having lower flux level. Here we confirm their finding with the larger {\it Swift} sample. We also see some potential (but opposite) trend with $\deltam$($B$) at $\lambda\lesssim2700$\,\AA, although it is less significant owing to the poorer SNR at shorter wavelengths.

We find trend between UV spectrum and host stellar mass, with SNe~Ia in more-massive galaxies tending to have lower flux levels at $\sim3000$--3500\,\AA. This trend is as expected given the tight correlation between SN decline rate and host stellar mass \citep[e.g.,][]{2014MNRAS.438.1391P}. A similar (but less significant) trend is also seen for the flux level below $3000$\,\AA. We find no clear trend between UV flux and host morphology.

The most intriguing result is probably from the relation with host metallicity. Here the split metallicity ($\rm12+\log(O/H)=8.6$) is selected to make approximately equally sized subgroups. We find a clear trend that SNe in more metal-rich galaxies (thus likely to have a metal-rich progenitor) tend to have lower flux levels at $\lambda<2700$\,\AA\ than their metal-poor counterparts, while the flux levels at longer wavelengths are similar for both subgroups. This is the first time we are able to see from observations that the host metallicity shows a correlation with SN~Ia UV spectra, and this metallicity effect apparently alters the flux exclusively at short wavelengths. We will discuss our findings in Section~\ref{sec:theory-compare}.

\subsection{UV flux ratio}
\label{sec:flux-ratio}
Following \citet{2016MNRAS.461.1308F}, we investigate the flux ratio of several spectral regions as a function of SN and host properties. In contrast to \citet{2016MNRAS.461.1308F}, here we are only able to examine two spectral regions owing to the smaller wavelength coverage of our {\it Swift} UV spectra; these correspond to the wavelength ranges of 2450--2620 and 2900--3150\,\AA. The flux ratio (labeled as $f_{2535}$ and $f_{3025}$, respectively) is then defined as the median flux of each region divided by the median flux for a wavelength range of 4000--4500\,\AA. The result is shown in Figure~\ref{flux-ratio}.

We see a trend between $f_{3025}$ and $\deltam$($B$), with larger $\deltam(B)$ having smaller $f_{3025}$. This is consistent with the mean spectra studied in Section~\ref{sec:mean-spectra}. Performing the linear fitting by using the Monte Carlo Markov Chain (MCMC) method \textsc{linmix} \citep{2007ApJ...665.1489K}, there is a $\sim99$\% probability that the slope is negative based on 10,000 MCMC realisations. We also derive a Pearson and Kendall correlation coefficient of $-0.52$ and $-0.35$, respectively. This trend still holds if we exclude two extremely fast decliners (with $\deltam(B)>1.6$\,mag) from the fitting. There is also some trend that SNe having lower $f_{3025}$ tend to be found in more massive galaxies, but this trend is less significant than that of $\deltam(B)$. No clear trend is found with other host properties, such as morphology and gas-phase metallicity. 

We next examine the trend with $f_{2535}$. We see a moderate trend with $\deltam(B)$ and host stellar mass, with SNe showing slower decline rates or found in more massive galaxies having smaller $f_{2535}$. No significant trend is found between $f_{2535}$ and host morphology. However, there is a $\sim2\sigma$ trend with host metallicity, in the sense that SNe in more metal-rich galaxies tend to have smaller $f_{2535}$. A $\sim95$\% probability of negative slope is determined, and a Pearson and Kendall correlation coefficient of $-0.60$ and $-0.27$, respectively. This result is also consistent with what we have seen using the mean spectra. The trend also holds if we further include the SN whose host was identified as a potential AGN in Section~\ref{sec:host}. The probability of a negative slope increases to 98\%, and the Pearson and Kendall correlation coefficients become $-0.66$ and $-0.32$, respectively. Excluding the SN hosted by the most metal-poor galaxy in our near-peak sample ($Z=8.267$) does not change the trend, but decreases its significance to the $\sim1\sigma$ level. In the future, a larger sample in this low-metallicity regime ($Z<8.5$) will be crucial to verify our results.

Our result is not sensitive to the uncertainty of the host-galaxy reddening or the choice of reddening laws. We performed a Monte Carlo experiment, where we assigned a random error in host $E(B-V)$ from 0 to 0.05\,mag for SNe without uncertainties measured in their host-galaxy reddening. A dispersion of only 0.003 ($\sim$2\%) is found for the slope of the $f_{2535}$ to host metallicity relation. This has a negligible effect on our result. We also compared our result with different reddening laws, such as the \citet{1999PASP..111...63F} law and two Large Magellanic Cloud reddening laws \citep{1999ApJ...515..128M}. The resulting differences in $f_{2535}$ are all $<1$\% and do not change the trends in Figure~\ref{flux-ratio}.

The results of the fitting and the correlation coefficients can be found in Table~\ref{uvflux-statistic}.

\begin{figure}
	\centering
		\includegraphics[scale=0.5]{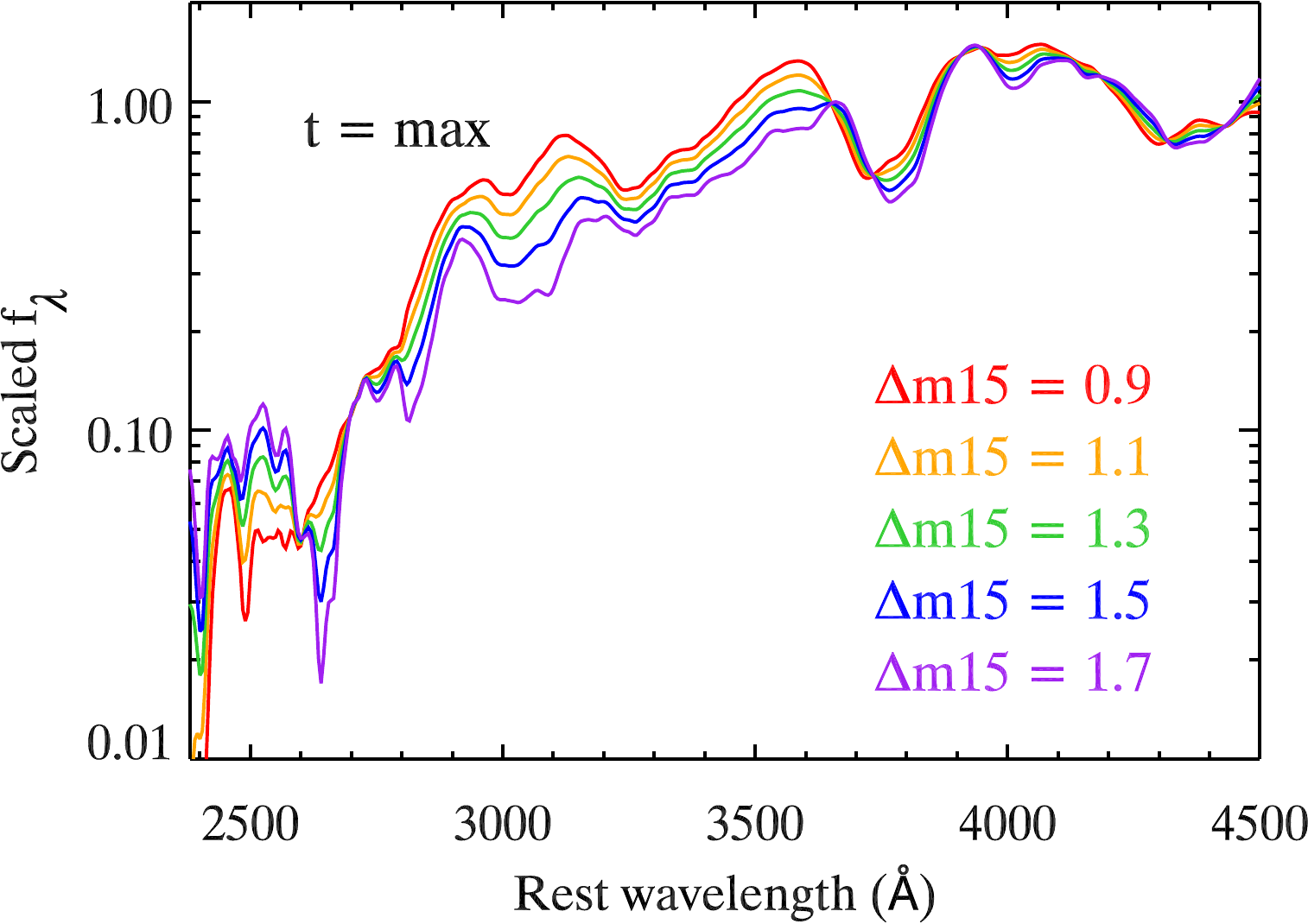}\\
		\vspace{0.25cm}
		\includegraphics[scale=0.5]{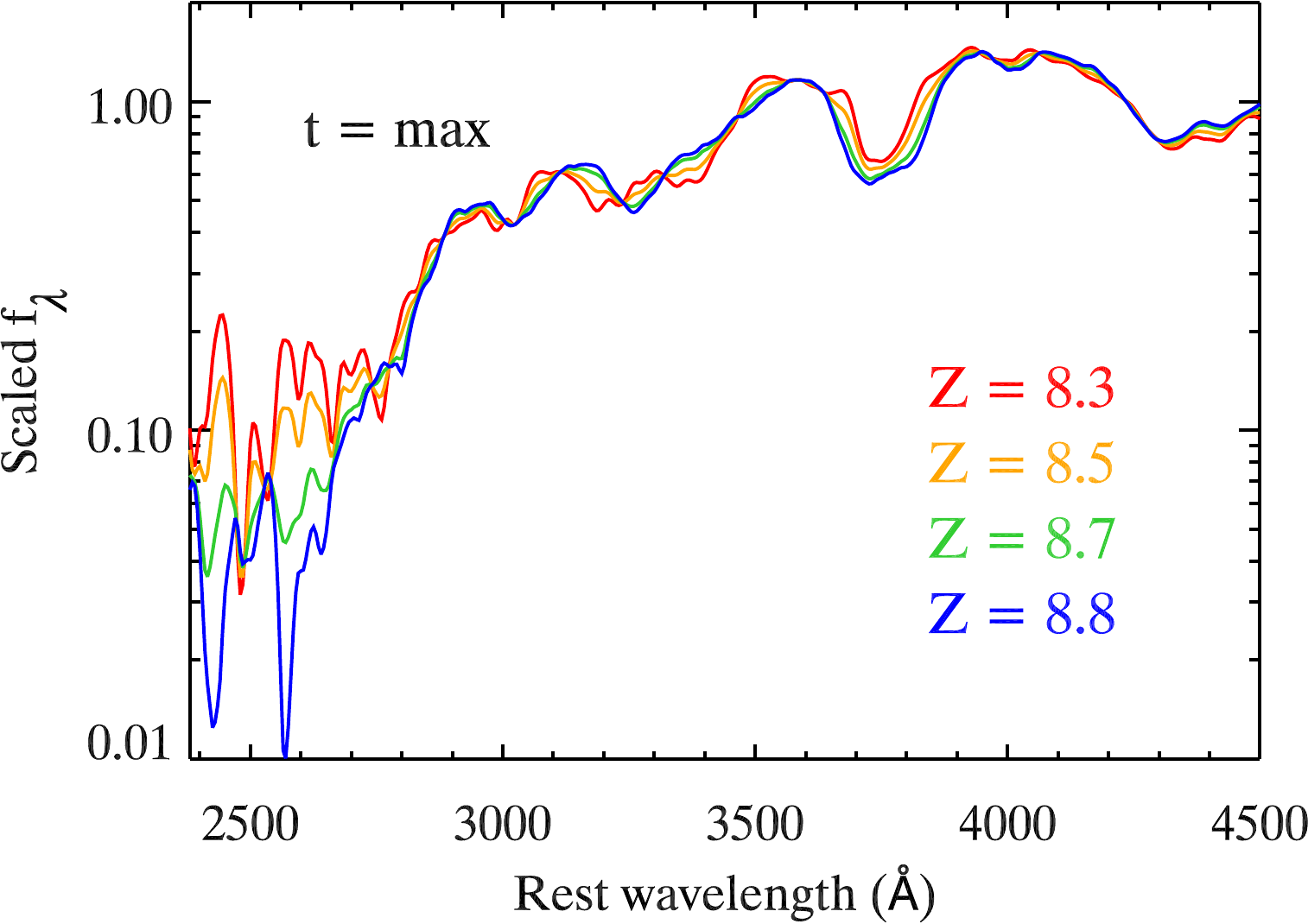}\\
                \caption{The data-driven templates created by SN~Ia UV near-peak spectra as a function of $\deltam(B)$ (upper panel) and host gas-phase metallicity $Z$ (lower panel).}
        \label{model-spec}
\end{figure}

\subsection{Data-driven spectral template}
\label{sec:model}
\begin{figure*}
	\centering
		\includegraphics[scale=0.73]{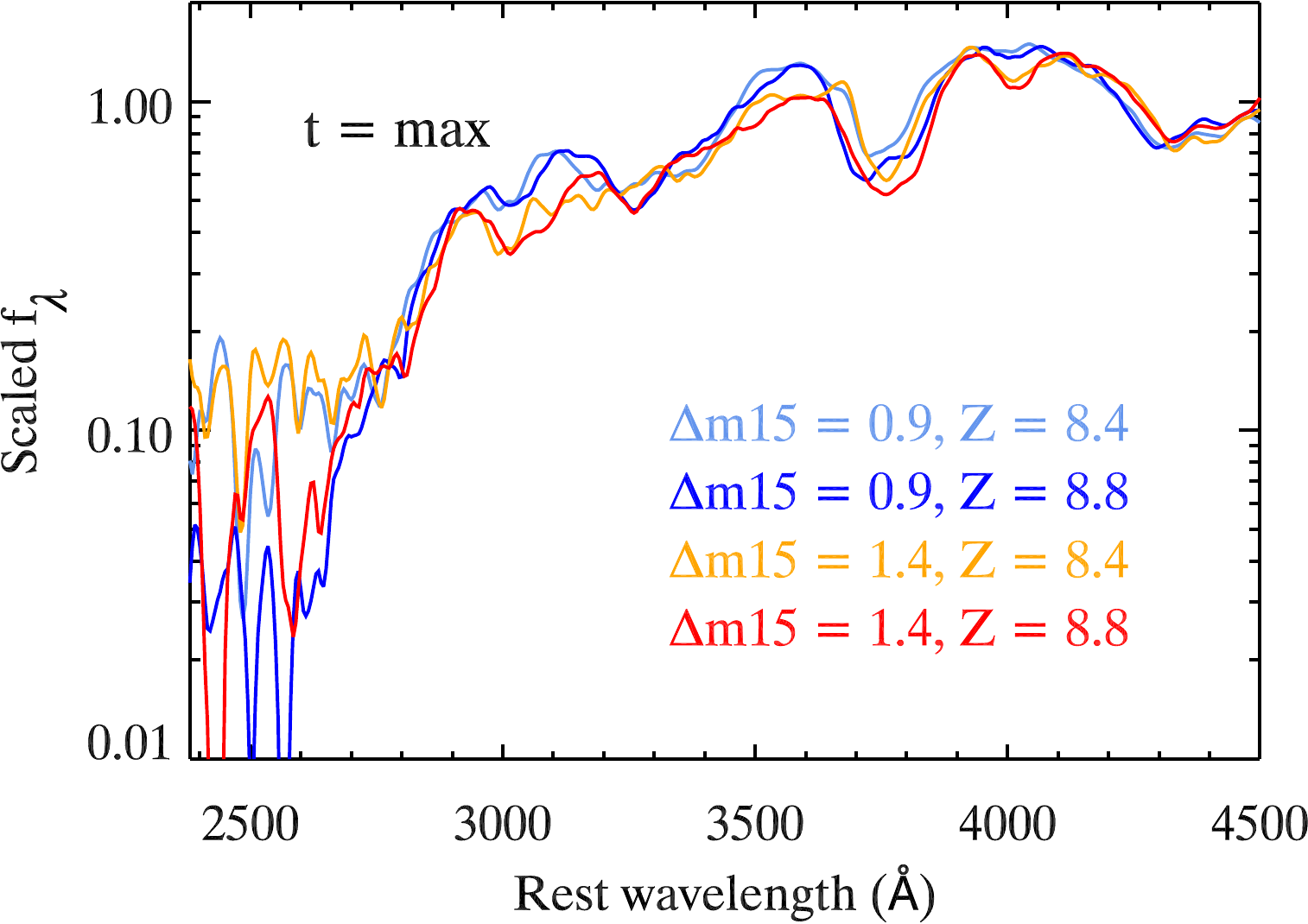}
                \caption{Same as Figure~\ref{model-spec}, but with both $\deltam(B)$ and $Z$ considered when fitting the template. Here we compare the templates generated by $\deltam(B)=0.9$ and 1.4\,mag, each with $Z=8.4$ and 8.8.}
        \label{model-spec2}
\end{figure*}

\begin{figure*}
	\centering
		\includegraphics[scale=0.79]{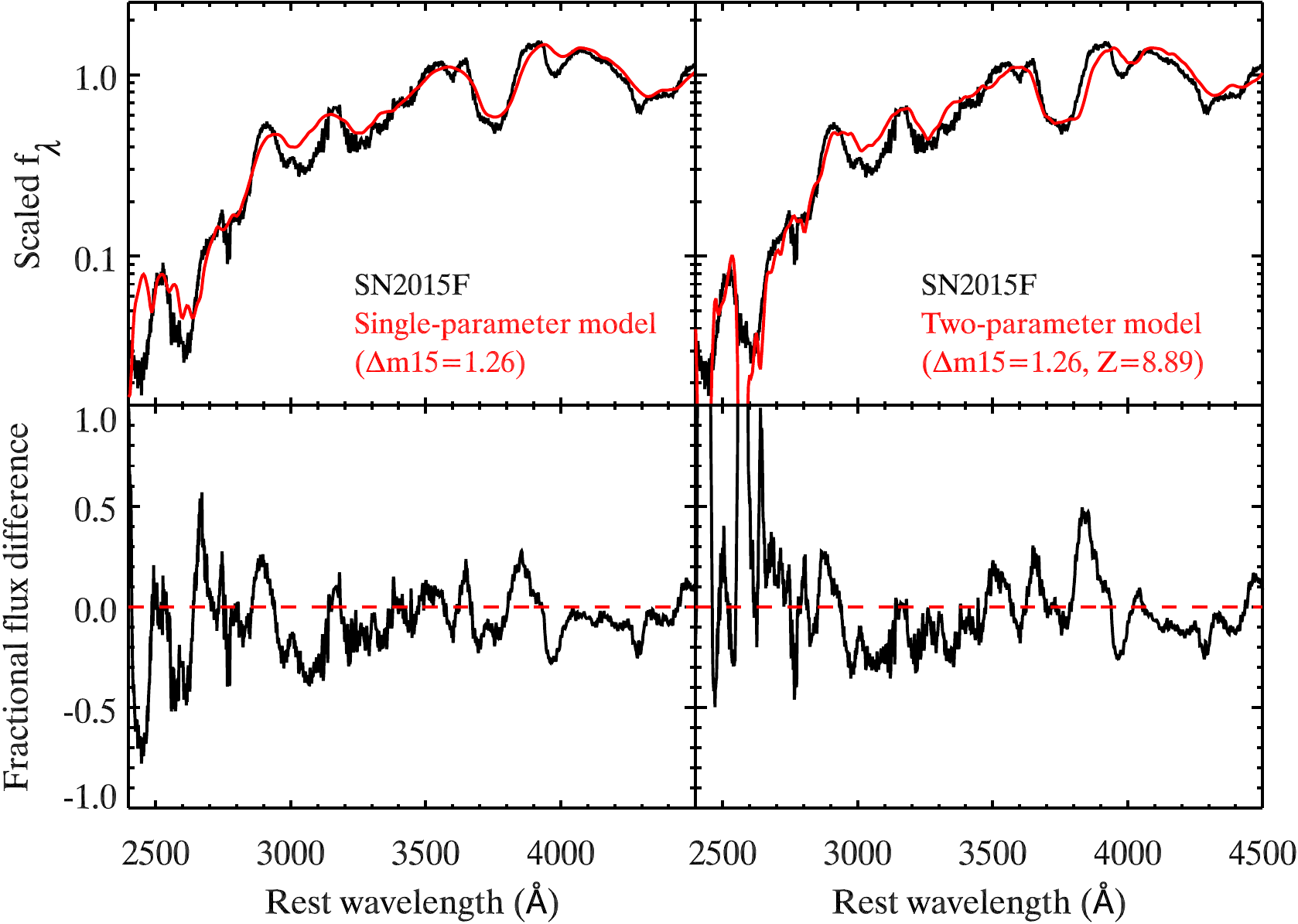}\\
                \caption{{\it Upper panels}: the left panel shows the comparison between the near-peak UV spectrum of SN~2015F and our template parameterised with $\deltam(B)$ only. The $\deltam(B)$ for SN~2015F \citep[1.26\,mag;][]{2015ApJS..221...22I} is used to create the template spectrum. The right panel is the same as the left panel, but with our UV template parameterised with both $\deltam(B)$ and host metallicity. Here we fix $\deltam(B)$ and vary only the metallicities to match the data (with only by-eye inspection). {\it Lower panels}: fractional flux differences from the template spectra shown in the upper panels.}
        \label{15F-model}
\end{figure*}

Here we generate the data-driven UV spectral template with our near-peak sample. We first create the template spectra with $\deltam(B)$ as the parameter. Following Eq.~(1) of \citet{2016MNRAS.461.1308F}, we generate a smoothed spectrum of each SN using an inverse-variance Gaussian filter \citep{2006AJ....131.1648B} and fit the smoothed flux for all spectra at each wavelength as a function of $\deltam(B)$. Note that the fitting performed here is unweighted, and we do not estimate the uncertainties for our template spectra. The result is shown in the upper panel of Figure~\ref{model-spec}.

The template spectra generally follow the same trend as we found in the data (e.g., the mean spectra). There is a strong correlation between the $\deltam(B)$ and the flux level at $\sim3000$--3500\,\AA. Similar to what was found by \citet{2016MNRAS.461.1308F}, our templates also show a pivot point at $\lambda\approx2700$\,\AA, where the templates tend to show an opposite trend blueward of 2700\,\AA\ in contrast to that redward of 2700\,\AA.

We next create the template spectra with host metallicity as the parameter. We fit the smoothed flux for all spectra in the sample at each wavelength as a function of host metallicity such that
\begin{equation}
\label{eq:model1}
f_{\lambda}=f_{8.6,\lambda}+s_{\lambda}\times(Z-8.6),
\end{equation}
where $f_{8.6,\lambda}$ represents the spectrum of a nominal SN~Ia of host gas-phase metallicity $Z=8.6$. The $s_{\lambda}$ is the deviation from that spectrum for a hypothetical SN~Ia of host $Z=9.6$. We present the parameters in Table~\ref{tab:model1}. The result is shown in the lower panel of Figure~\ref{model-spec}.

There is an obvious trend that the metallicity has a substantial effect in altering the flux level of our template spectra, and is particularly significant at $\lambda\lesssim2700$\,\AA. This implies that the near-peak UV spectra (at least for $\lambda\lesssim2700$\,\AA) could be promising indicators of SN~Ia progenitor metallicity.

As noted above, it is apparent that $\deltam(B)$ and metallicity are both affecting the UV spectra of SNe~Ia (although likely in different ways). To disentangle this degeneracy, we try to construct UV templates based on both $\deltam(B)$ and metallicity parameters to explore the two-dimensional parameter space. The formula we use for the fitting is 
\begin{equation}
\label{eq:model2}
f_{\lambda}=f_{(1.1,8.6),\lambda}+s_{1,\lambda}\times(\deltam(B)-1.1)+s_{2,\lambda}\times(Z-8.6),
\end{equation}
where $f_{1.1,8.6,\lambda}$ represents the spectrum of a nominal $\deltam(B)=1.1$\,mag SN~Ia with the host galaxy having a gas-phase metallicity $Z=8.6$. Also, $s_{1,\lambda}$ is the deviation from that spectrum for a hypothetical SN~Ia of $\deltam(B)=2.1$\,mag and host $Z=8.6$, while $s_{2,\lambda}$ is the deviation from the same spectrum for a hypothetical SN~Ia of $\deltam(B)=1.1$\,mag and host $Z=9.6$. We present the parameters in Table~\ref{tab:model2}. The result is shown in Figure~\ref{model-spec2}. Our two-parameter templates clearly indicate that $\deltam(B)$ and metallicity have distinct effects on the UV spectrum.

Figure~\ref{15F-model} shows the comparison between the UV spectrum of SN~2015F and our templates. The templates parameterised with only $\deltam(B)$ and both $\deltam(B)$ and host metallicity are compared. Since we do not have the host metallicity measured for SN~2015F, we fix $\deltam(B)$ to that of SN~2015F (which is already known from the light-curve fitting) and only let the metallicity vary.  Note that the comparison made here is simply by-eye inspection; we did not fit our templates to the data.

It is clear that with the UV template having only a single parameterisation in light-curve shape, it is difficult to interpret the whole UV spectrum (especially for the behaviour at short wavelengths). The template which takes the metallicity into consideration is apparently more consistent with the data. By comparing the UV spectrum with our templates, we find that SN~2015F is likely to reside in a metal-rich host environment. This is probably not surprising given its relatively massive host galaxy ($\log\,(M_{\rm stellar}/{\rm M}_{\odot})=10.56$). Thus, it implies that SN~2015F could have a high-metallicity progenitor.

However, one should be cautious in using these templates. Firstly, we have only a relatively small number of SNe with measured host metallicities in certain regions of parameter space (especially those SNe at both ends of the $\deltam(B)$ distribution of P18). A discrepancy between the host metallicity measured in this work and SN progenitor metallicity is also expected, but the relative trend existing in both mean spectra and spectral templates should be real. Secondly, the real mechanisms forming the UV spectra of SNe~Ia should be far more complicated than a simple description of a two-parameter template \citep[e.g.,][]{2000A&A...363..705M}. Other parameters (from the progenitor and/or explosion) could also play a role, but the strong dependence of SN~Ia UV spectra on light-curve shape and metallicity implies that they are likely important factors.

\begin{table}
\center
\caption{UV spectral template parameters for Eq.~(\ref{eq:model1}). The entire table is available online as supplementary material. A portion is shown here for guidance regarding its form and content.}
\begin{tabular}{lcc}
\hline\hline
Wavelength & $f_{8.6,\lambda}$  & $s_{\lambda}$\\
(\AA)      &                         &                           \\
\hline
2350  &   0.0757  &   $-0.1038$\\
2355  &   0.0887  &   $-0.1526$\\
2360  &   0.0926  &   $-0.1939$\\
2365  &   0.0938  &   $-0.2063$\\
2370  &   0.0934  &   $-0.1899$\\
2375  &   0.0885  &   $-0.1424$\\
2380  &   0.0802  &   $-0.0711$\\
2385  &   0.0737  &   $-0.0215$\\
2390  &   0.0716  &   $-0.0200$\\
2395  &   0.0700  &   $-0.0624$\\
\hline
\label{tab:model1}
\end{tabular}
\end{table}

\begin{table}
\center
\caption{The same as Table~\ref{tab:model1}, but for parameters in Eq.~(\ref{eq:model2}).}
\begin{tabular}{lccc}
\hline\hline
Wavelength & $f_{(1.1,8.6),\lambda}$ & $s_{1,\lambda}$ & $s_{2,\lambda}$\\
(\AA)      &                    &             & \\
\hline
2350  &    0.1431  &    0.2148  &   $-0.1363$\\
2355  &    0.1587  &    0.2460  &   $-0.2306$\\
2360  &    0.1625  &    0.2628  &   $-0.3171$\\
2365  &    0.1628  &    0.2836  &   $-0.3761$\\
2370  &    0.1621  &    0.3000  &   $-0.3994$\\
2375  &    0.1553  &    0.2941  &   $-0.3798$\\
2380  &    0.1443  &    0.2681  &   $-0.3234$\\
2385  &    0.1342  &    0.2264  &   $-0.2700$\\
2390  &    0.1261  &    0.1714  &   $-0.2405$\\
2395  &    0.1171  &    0.1160  &   $-0.2364$\\
\hline
\label{tab:model2}
\end{tabular}
\end{table}

\subsection{Spectral evolution}
\label{sec:evolution}
\begin{figure*}
	\centering
		\includegraphics[scale=0.35]{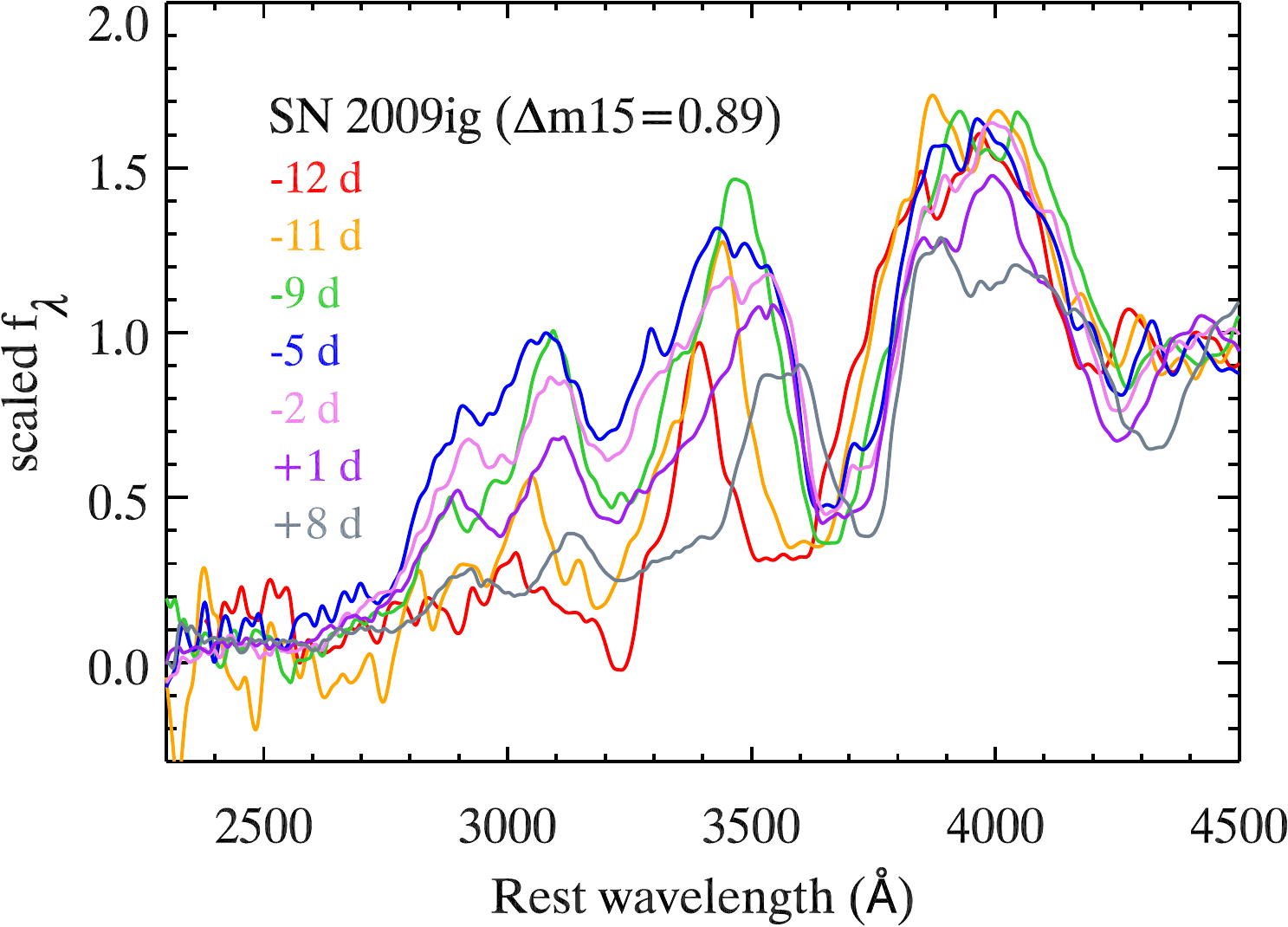}
		\includegraphics[scale=0.35]{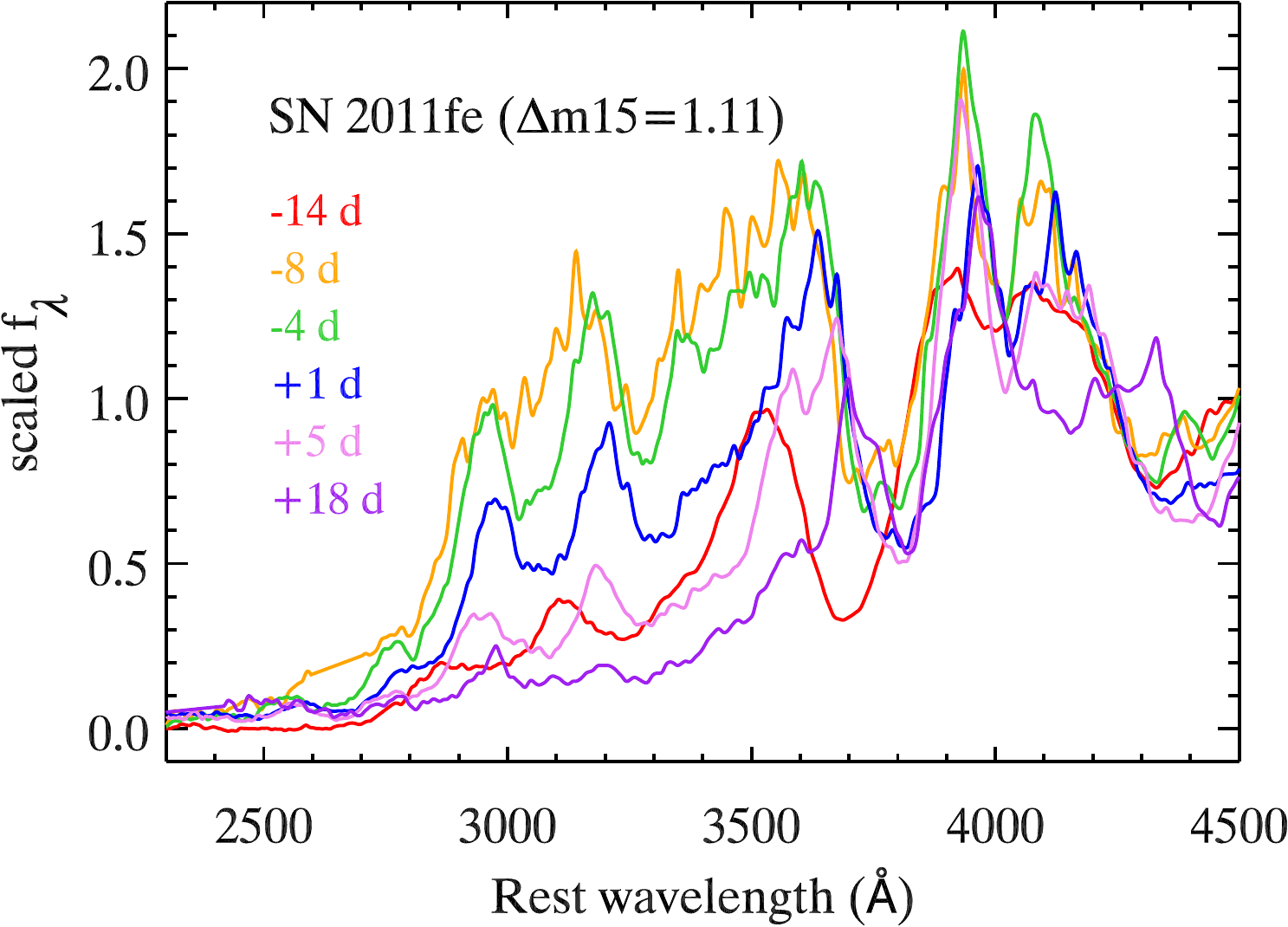}
		\includegraphics[scale=0.35]{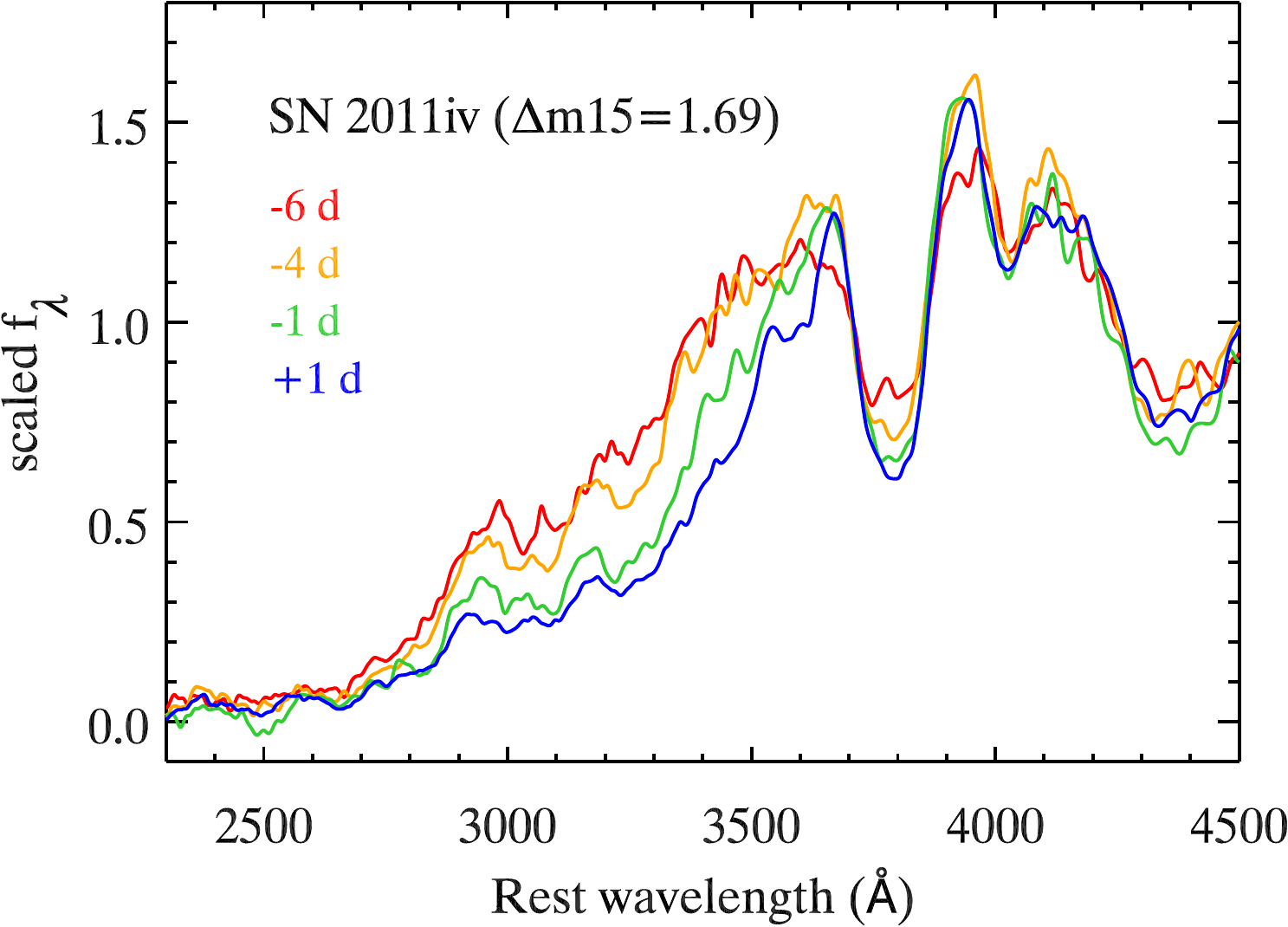}
                \caption{The UV spectral sequence of SN~2009ig (left), SN~2011fe (middle), and SN~2011iv (right). The phases in each panel are relative to $B$-band maximum light.}
        \label{uv-evolution}
\end{figure*}

Many SNe in our sample have a well-observed spectral sequence, making them ideal targets for studying the temporal evolution of UV spectral features. 

In Figure~\ref{uv-evolution}, we present examples of UV spectral sequences for SNe with several different values of $\deltam(B)$. The epochs of our spectral observations span $\sim\pm1$--2 weeks relative to the time of peak brightness. There is clear evolution for the UV feature at $\lambda\approx3000$--3500\,\AA. The flux level rises dramatically (relative to the flux at $\sim4000$\,\AA) with time after the SN explosion and peaks $\sim1$ week before optical maximum light. It then decreases slowly (relative to the rising) with time after peak brightness. This trend is similar for both low- and high-$\deltam(B)$ SNe, although the high-$\deltam(B)$ SNe generally exhibit much weaker (and more featureless) UV spectra than do low-$\deltam(B)$ SNe at similar phases.

\begin{figure}
	\centering
		\includegraphics[scale=0.5]{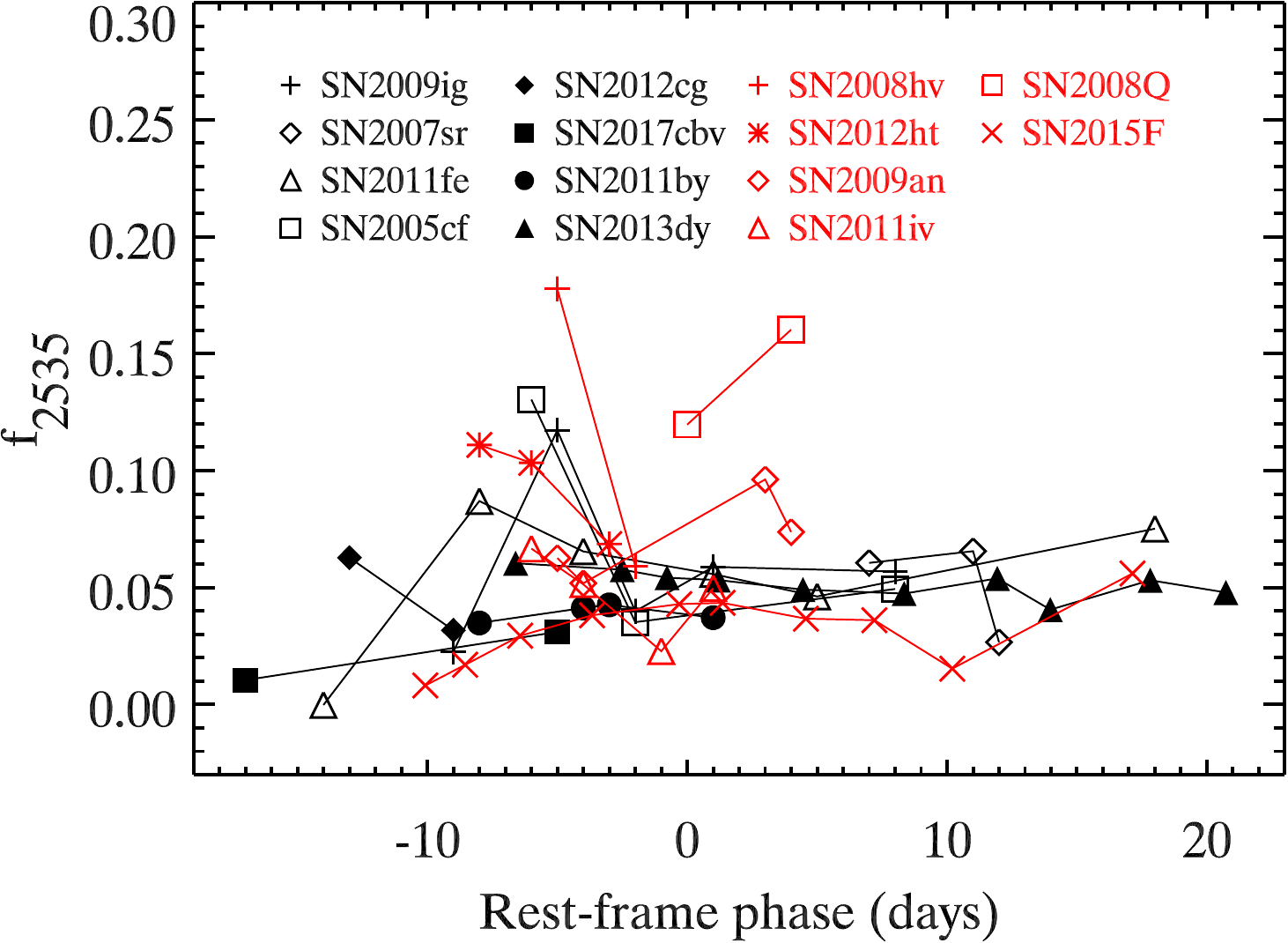}\\
		\includegraphics[scale=0.5]{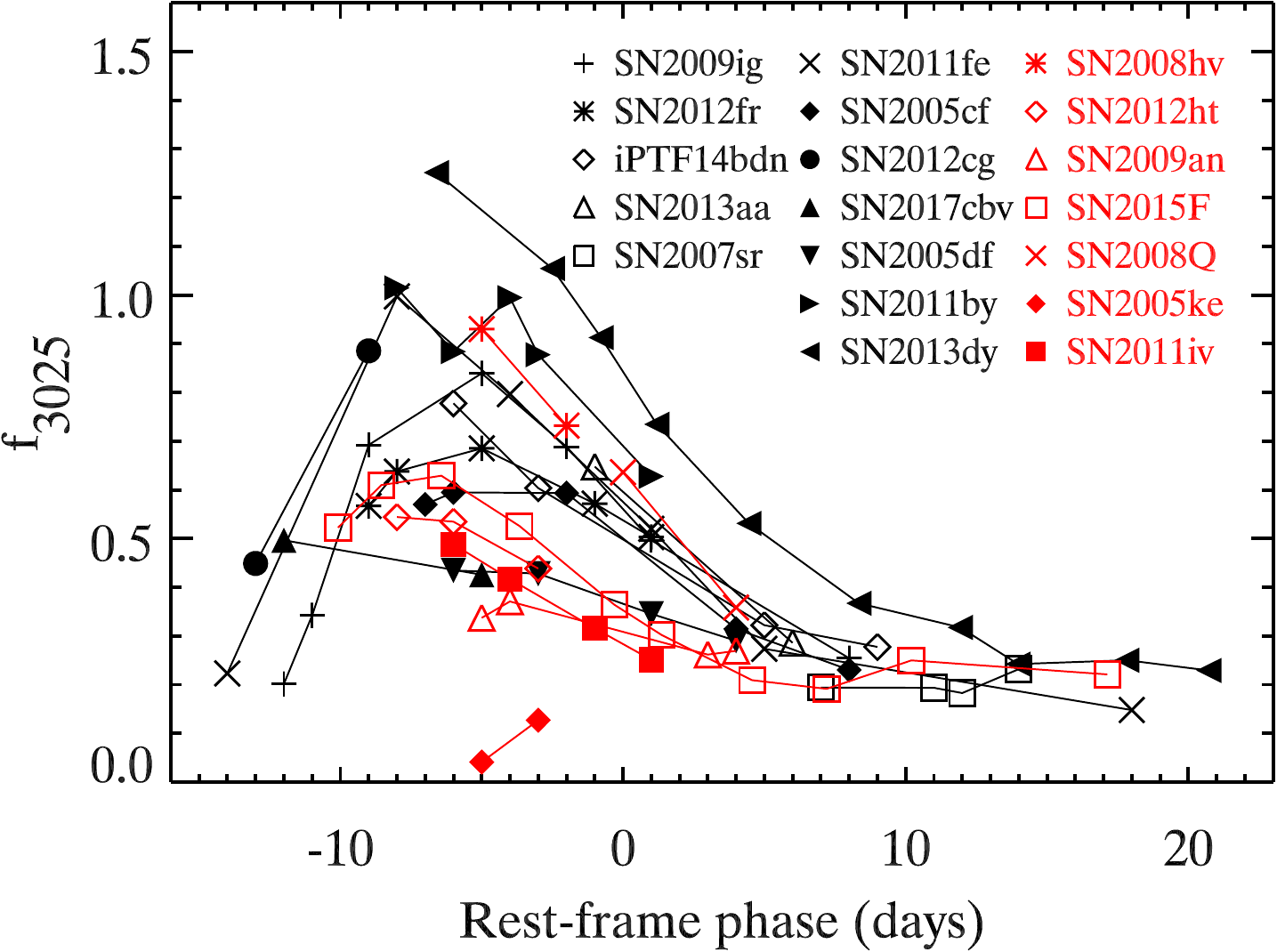}\\
                \caption{The temporal evolution of the UV flux ratios $f_{2535}$ and $f_{3025}$. The black and red symbols correspond to the SNe with $\deltam(B)<1.2$\,mag and $\deltam(B)>1.2$\,mag, respectively.}
        \label{uv-evolution2}
\end{figure}

We examine the temporal evolution of the flux ratios $f_{2535}$ and $f_{3025}$ in Figure~\ref{uv-evolution2}. Similar to what we have seen in Figure~\ref{uv-evolution}, there is strong evolution of $f_{3025}$ with time. The feature also peaks $\sim1$ week before optical maximum light, indicating that early-time observations are more critical for the UV than for the optical. However, we do not see clear evolution for $f_{2535}$. 

By examining each individual SN, we find that SN~2008hv and SN~2008Q are apparent outliers to these trends (in terms of their $\deltam(B)$). They are noticeably bright in the UV despite their relatively fast decline rates ($\deltam(B)=1.25$ and 1.40\,mag, respectively). We note that both SNe were classified as ``NUV-blue'' SNe~Ia by \citet{2013ApJ...779...23M}. This represents a relatively minor group of SNe~Ia (at least at low redshifts) which shows distinctly high NUV to optical ratios. SN~2011fe and SN~2011by were also classified as NUV-blue SNe~Ia. It is clear that their peak flux ratios $f_{3025}$ are higher than the majority of our sample. SN~2013dy and SN~2012cg are also similar to those NUV-blue events as described by \citet{2013ApJ...779...23M}. Intriguingly, we find that these NUV-blue SNe~Ia are not necessarily blue at shorter wavelength (e.g., $f_{2535}$). This implies that the UV spectrum at those wavelengths might be influenced by different physics (e.g., progenitor metallicity).

\section{Discussion}
\label{sec:discussion}
\subsection{Comparison with previous studies}
\label{sec:theory-compare}
\citet{2000ApJ...530..966L} modeled the effects of progenitor metallicity on SNe~Ia by modifying the metallicity of the unburned C+O layer in the W7 model of \citet{1984ApJ...286..644N}. They produced synthetic spectra of a variety of progenitor metallicities with wavelengths extended to the UV. They found that the UV continuum of the synthetic spectra is strongly affected by the metallicity in the unburned layer, with higher metallicities having lower flux levels. This trend is likely caused by the stronger UV line blanketing with increasing IGEs in the SN outer layers. Higher metallicity also results in a shallower line-forming region and causes the UV features to move blueward in the spectra. 

Using a different spectral synthesis code, \citet{2008MNRAS.391.1605S} also showed that the variation of IGE abundance does not significantly affect the optical, but is significant in the UV part of the spectrum. However, they found that the UV flux level may either increase or decrease with increasing abundance of heavy elements, depending on the models they used. \citet{2012MNRAS.427..103W} adopted the radiative transfer codes based on \citet{2008MNRAS.391.1605S} but different explosion models to investigate the effect of metal content in the SN outer layers on the UV spectrum. They concluded that the flux at $\sim3000$--3500\,\AA\ relative to that near 4000\,\AA\ is a good tracer of metallicity.

In this work, we use the host-galaxy metallicity as a proxy for progenitor metallicity to examine the correlation with SN~Ia UV spectra. Our result is generally consistent with the predictions of \citet{2000ApJ...530..966L}. Their model spectra at 15~days after explosion (i.e., similar to the phase studied in this work) also reveal that the flux levels start to show increasingly large differences at $\lambda\lesssim2700$\,\AA\ between progenitor metallicities. This is consistent with both the mean spectra and data-driven spectral templates in this work. We do not see a significant trend between flux level at $\sim3000$--3500\,\AA\ and metallicity, which contradicts the model predictions of \citet{2012MNRAS.427..103W}.

\begin{figure}
	\centering
		\includegraphics[scale=0.49]{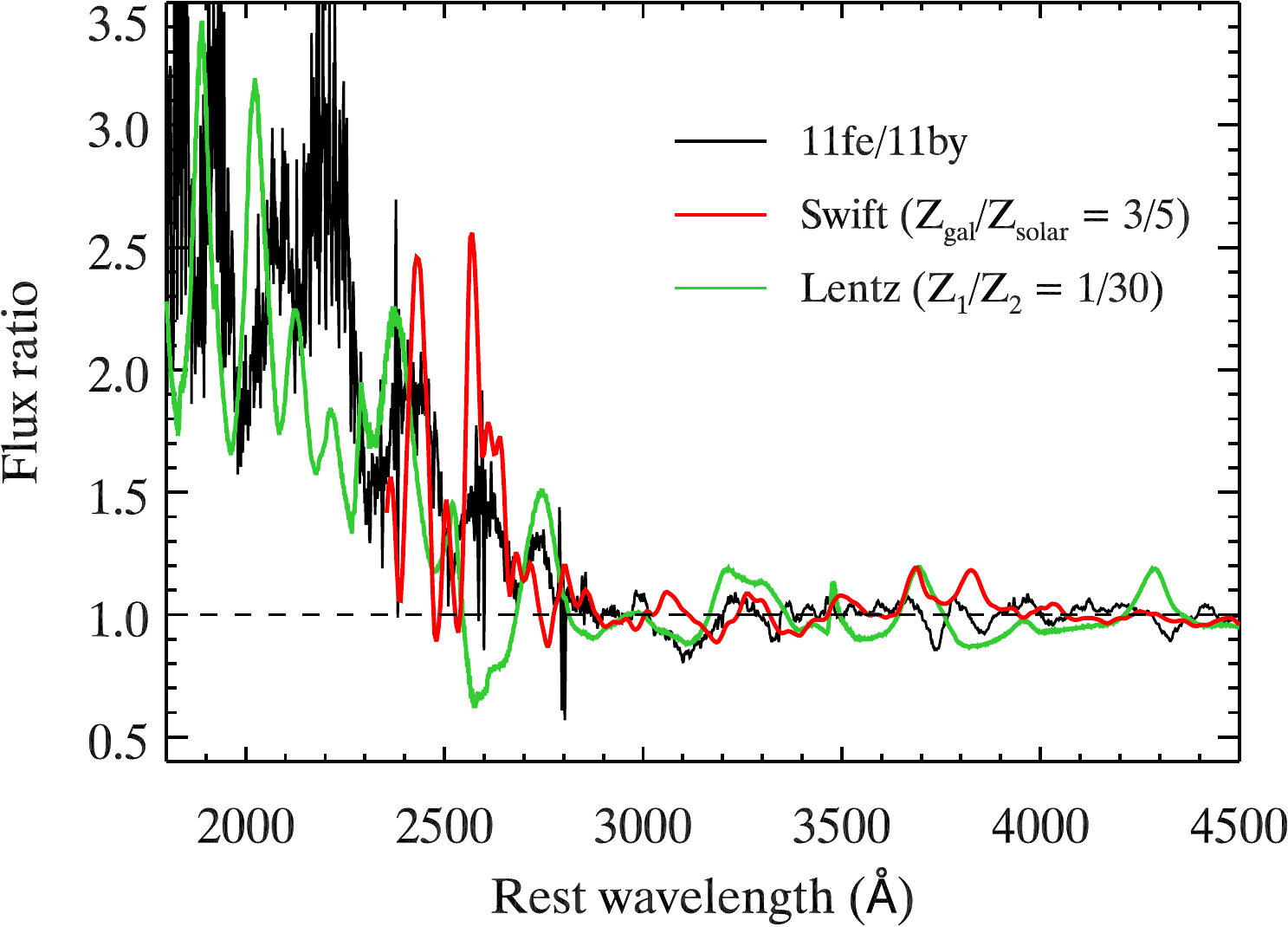}
                \caption{The flux-ratio spectrum of SN~2011fe to SN~2011by observed by {\it HST}/STIS (in black). The red line represents the ratio of {\it Swift} template spectra of SNe~Ia in 0.6\,$\rm Z_{\odot}$ host galaxies relative to those in solar-metallicity host galaxies. The green line is the ratio of model spectra from \citet{2000ApJ...530..966L} with progenitor metallicity ratios of 1/30.}
        \label{uv-model-compare}
\end{figure}

\citet{2013ApJ...769L...1F} studied the ``twin" SN~2011by and SN~2011fe and found some evidence of a metallicity effect on the SN~Ia UV spectrum. These two SNe have nearly identical optical spectra and light curves but very different UV spectra. They compared the observed difference to those of model spectra in \citet{2000ApJ...530..966L} and concluded that the difference in their progenitor metallicities is likely the main source driving the discrepancy in the UV (without affecting the optical). Following their methods, we produce the near-peak flux-ratio spectrum of SN~2011fe to SN~2011by and present it in Figure~\ref{uv-model-compare}. By using flux ratios, we can avoid the differences of spectral features between models and data, and focus on the differences between models (e.g., the metallicity). We then compare the flux-ratio spectrum of the twin SNe to that of our data-driven templates as parameterised by the host metallicity. We find that SN~2011by is likely to have higher progenitor metallicity than that of SN~2011fe. For example, the flux ratio produced by 0.6\,$\rm Z_{\odot}$ and 1.0\,$\rm Z_{\odot}$ \citep[using the solar value of 8.69;][]{2001ApJ...556L..63A} template spectra could match the data quite well (the comparison made here is simply a by-eye inspection). The model flux-ratio spectrum of metallicity factor $Z_{1}/Z_{2}=1/30$ from \citet{2000ApJ...530..966L} is also overplotted for comparison.

Our results support the previous findings that SN~2011by and SN~2011fe could have solar (or above solar) and subsolar progenitor metallicities, respectively. Nonetheless, the ``direct'' interpretation or measurement of the progenitor metallicity using our templates could be problematic owing to a potential discrepancy between the host-galaxy and SN-progenitor metallicities. The metallicity measured close to the position of the progenitor system (e.g., determined from the local environment) should be considered to build more precise templates.

\subsection{Implications for cosmology}
\label{sec:cosmology}
\begin{figure}
	\centering
		\includegraphics[scale=0.5]{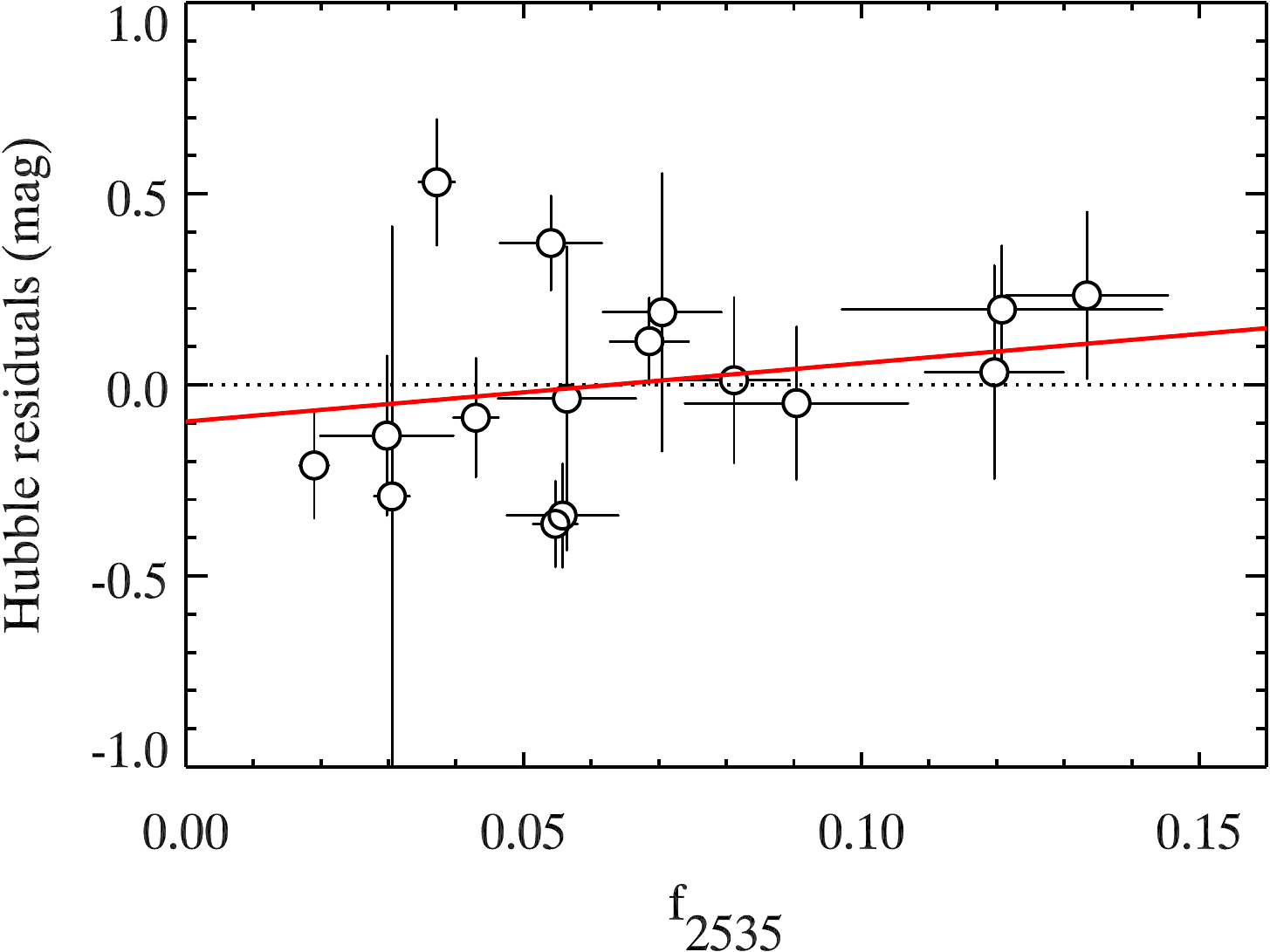}\\
		\vspace{0.25cm}
		\includegraphics[scale=0.5]{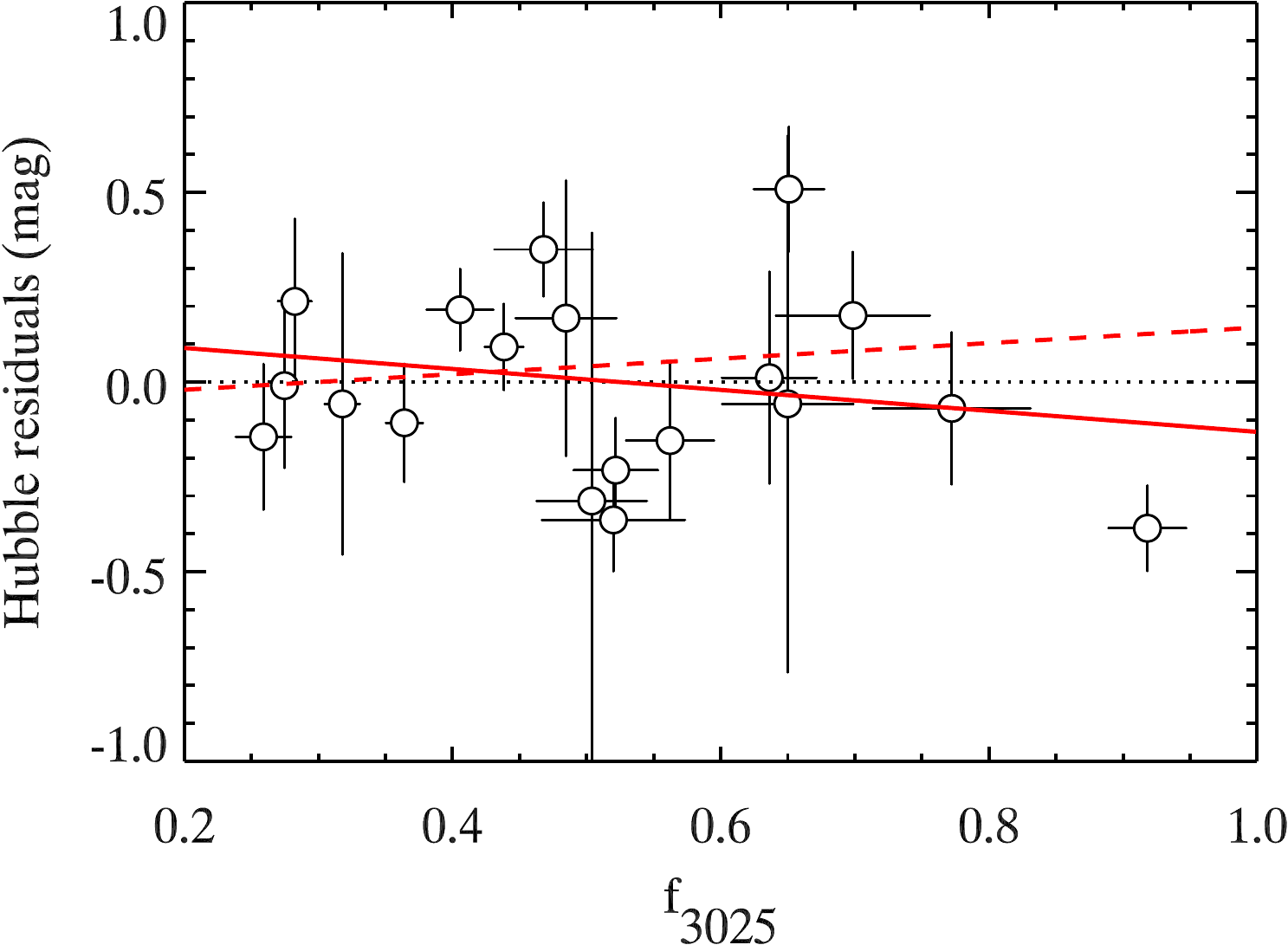}\\
		\vspace{0.25cm}
		\includegraphics[scale=0.49]{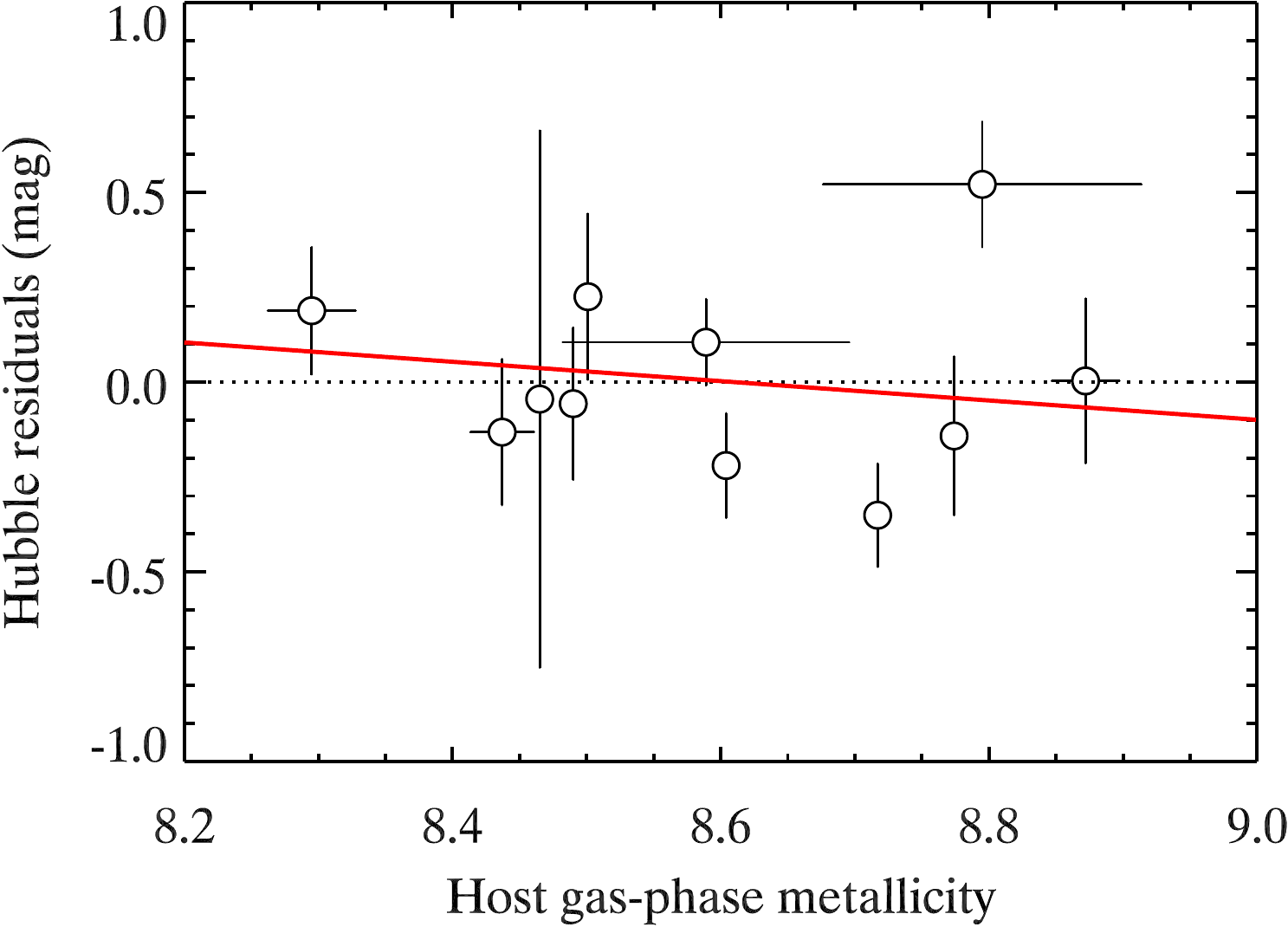}
                \caption{The Hubble residuals from our near-peak sample as a function of flux ratios $f_{2535}$ (top) and $f_{3025}$ (middle), and host gas-phase metallicity (bottom). The red line in each panel represents the linear fit to the data. The dashed line in the middle panel shows the same linear fitting, but excluding the SN of highest $f_{3025}$ in our sample, which causes a significant trend with Hubble residual that is otherwise not present. The dotted line in each panel represents the line of zero residual.}
        \label{uvflux-hrs}
\end{figure}

Theory predicts that the amount of $^{56}$Ni produced during the SN~Ia explosion shows a strong dependence on the progenitor metallicity \citep[e.g.,][]{2003ApJ...590L..83T}. More metal-rich progenitors will likely generate more stable and neutron-rich $^{58}$Ni instead of the radioactive $^{56}$Ni, and therefore result in fainter SNe~Ia. Host-galaxy observations also suggested that SNe found in more metal-rich environments tend to be fainter than those in metal-poor environments \citep[e.g.,][]{2011ApJ...743..172D,2013ApJ...770..108C,2014MNRAS.438.1391P}. This dependence is significant even after the empirical corrections for SN light-curve width and colour. Given the underlying metallicity evolution with redshift \citep[e.g., $\sim0.2$\,dex from $z=0$ to 1;][]{2005ApJ...635..260S}, this could introduce substantial bias in the current cosmological analyses.

Theoretically, the UV spectrum of a SN~Ia should be a more direct probe of progenitor metallicity than the host galaxy. Accordingly, we inspect the correlation between the UV flux ratio and SN luminosity as parameterised by Hubble residuals to see if it shows a stronger trend than that with host galaxies. The Hubble residuals are defined as the difference between the observed rest-frame $B$-band SN apparent magnitude corrected for light-curve width and colour, and the peak SN magnitude expected in our assumed cosmological model. We use \textsc{salt2} \citep{2007A&A...466...11G} to fit the SN light curves, and we adopt similar procedures and cuts on light-curve width and colour as those described by \citet{2018MNRAS.475..193F}. However, we did not apply the correction for the host mass step as we are actually searching for potential variables which could have the same effect as host stellar mass (e.g., metallicity). For the 9 SNe having precise distances measured by \citet{2016ApJ...826...56R}, we use their Cepheid-based distance modulus instead of that derived from the cosmological models. For those without Cepheid distances, the distance modulus is determined by the \textsc{python} script \textsc{cosmos} and assumes $\rm H_{0}=67.7$\,km\,s$^{-1}$\,Mpc$^{-1}$ and a flat universe with $\Omega_{M}=0.3$.

The result is shown in Figure~\ref{uvflux-hrs}. Previous studies revealed that SNe in more metal-rich environments tend to have more negative Hubble residuals than their metal-poor counterparts \citep[e.g.,][]{2014MNRAS.438.1391P}. If the host metallicity traces the progenitor metallicity well, we should expect SNe with lower flux ratio $f_{2535}$ (caused by higher progenitor metallicity) to have more negative Hubble residuals. However, we do not see a significant trend with $f_{2535}$; if anything, SNe with lower $f_{2535}$ tend to have more negative Hubble residuals. There is a $\sim72$\% probability that the slope is positive based on 10,000 MCMC realisations. In comparison, we see an opposite trend with $f_{3025}$. We find this trend is likely caused by an outlier in the sample (see the middle panel of Figure~\ref{uvflux-hrs}), and which otherwise is not present. We also show the Hubble residuals as a function of host gas-phase metallicities (the bottom panel of Figure~\ref{uvflux-hrs}). However, we are unable to reproduce the significant trend found by previous host studies, although it is consistent with the absence of a trend we found with $f_{2535}$.

It is likely that our results suffer from large uncertainties, given that our sample is relatively small and nearby. At $z\lesssim0.01$, precise distance measurement becomes extremely difficult owing to peculiar velocities, and thus those Hubble residuals only very weakly depend on the cosmological model. It would be interesting to re-examine these relations by extending the current analysis to a higher-$z$ SN sample in the future.

\subsection{Early-time UV spectra}
\label{sec:early}
\begin{figure}
	\centering
	    \includegraphics[scale=0.47]{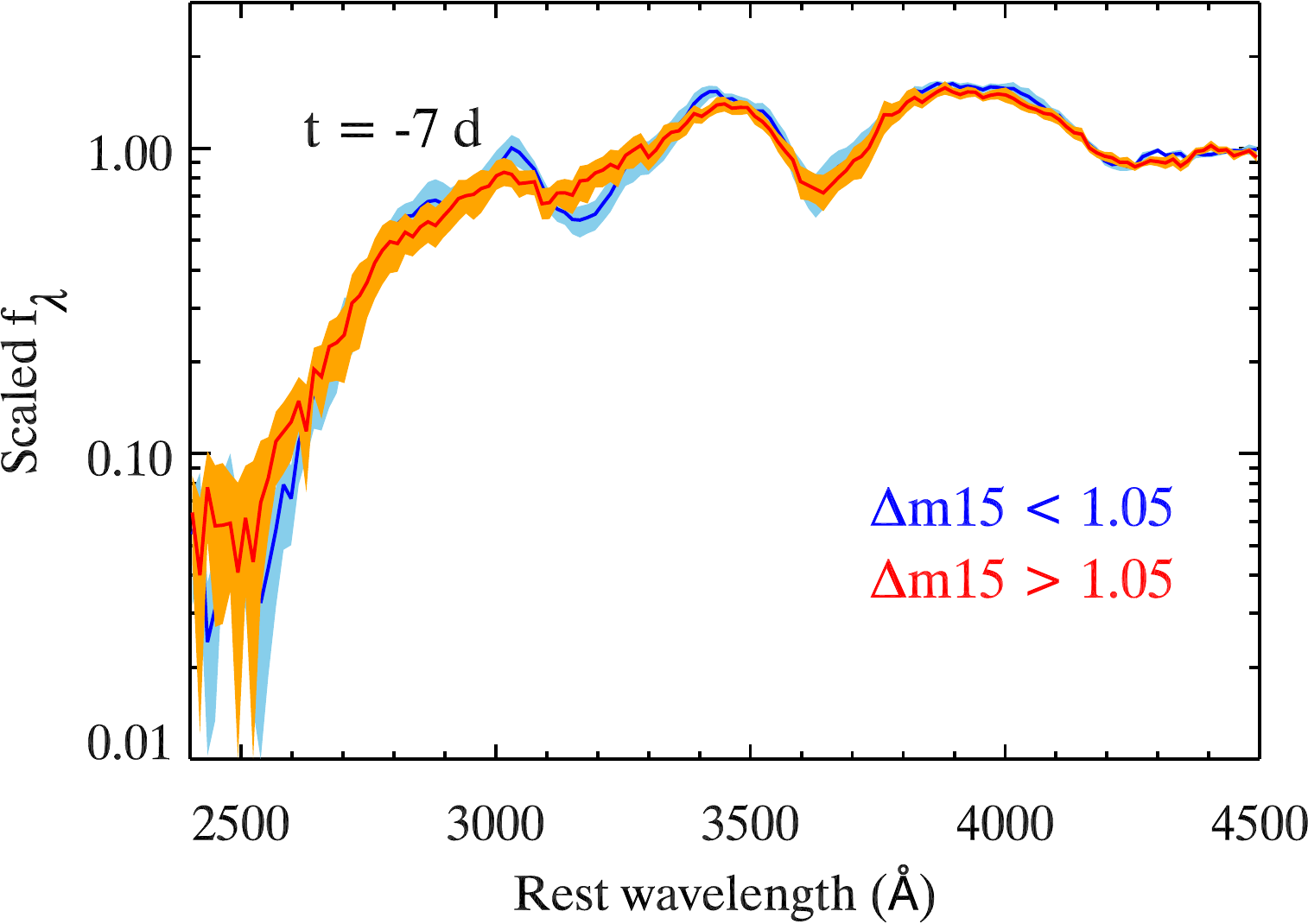}\\
		\vspace{0.25cm}
		\includegraphics[scale=0.47]{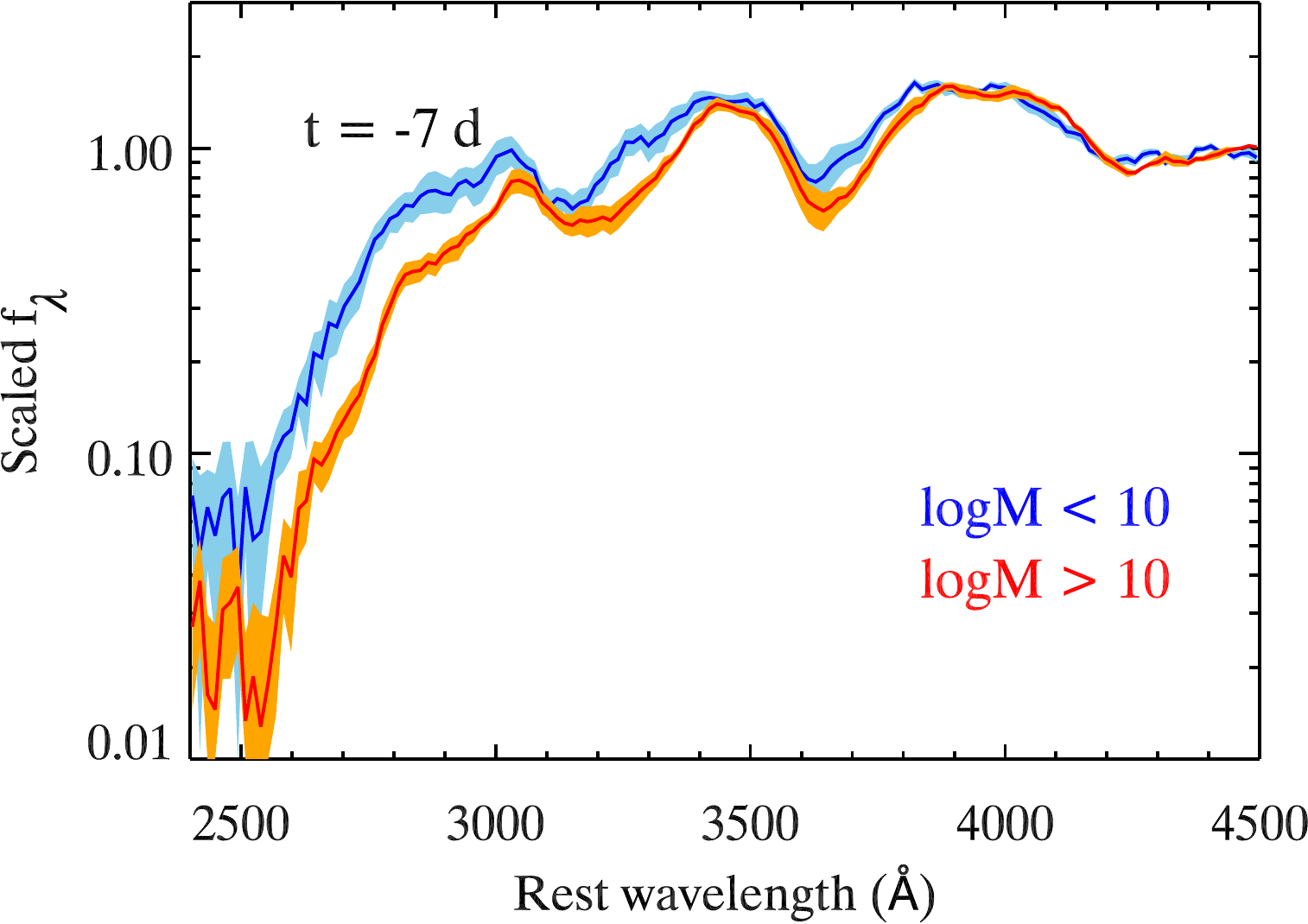}\\
		\vspace{0.25cm}
		\includegraphics[scale=0.47]{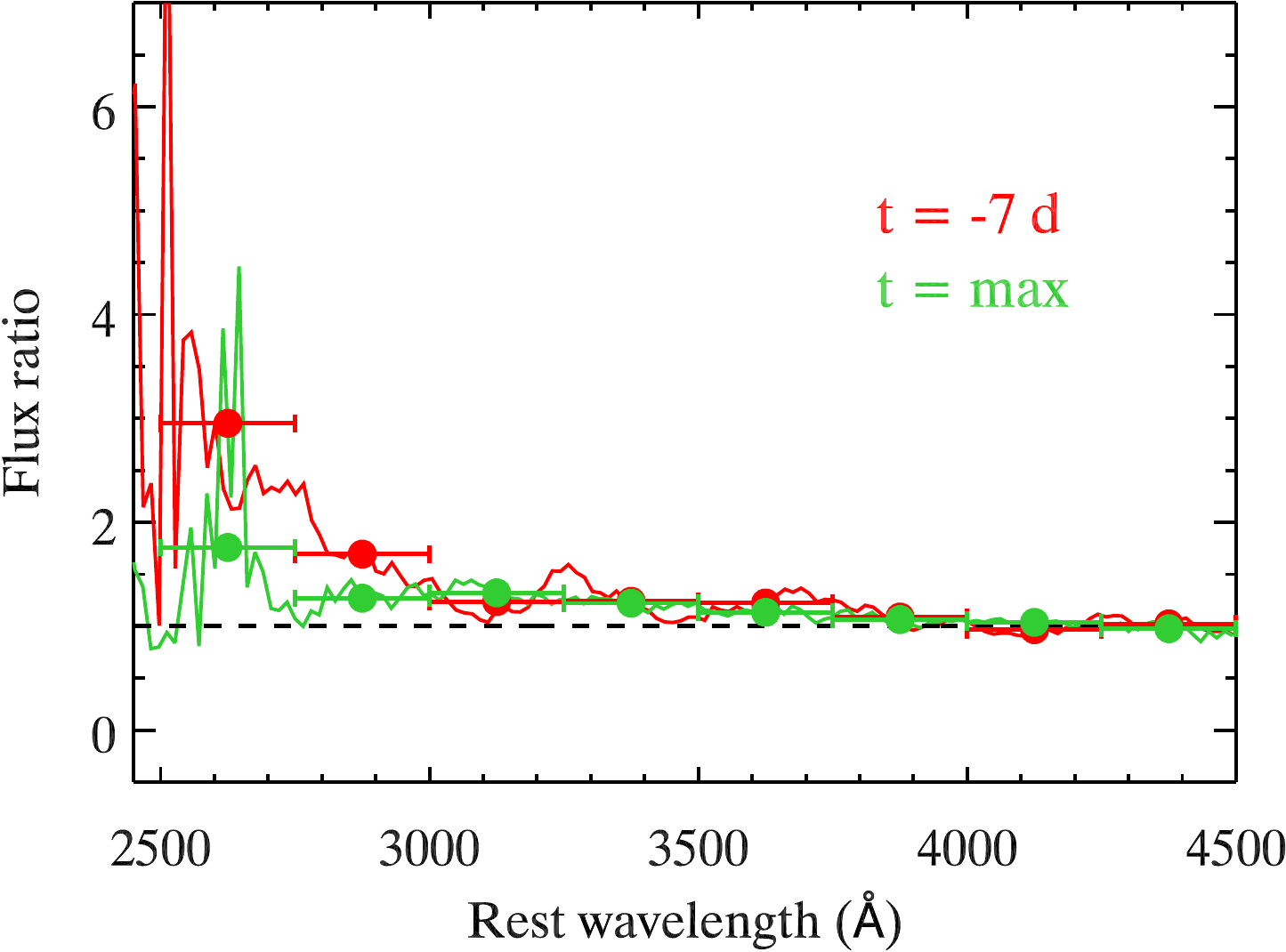}
                \caption{{\it Upper}: The mean spectra of the {\it Swift} early-time sample ($t \approx -7$\,days) based on $\deltam(B)$. The mean spectra of the slower decliners ($\deltam(B)<1.05$\,mag) and faster decliners ($\deltam(B)>1.05$\,mag) are shown with blue and red lines, respectively. {\it Middle}: Same as the upper panel, but with the host-galaxy stellar mass instead. The mean spectra of SNe in low-mass host galaxies ($\log\,(M_{\rm stellar}/{\rm M}_{\odot})<10$) and high-mass galaxies ($\log\,(M_{\rm stellar}/{\rm M}_{\odot})>10$) are represented with blue and red lines, respectively. {\it Bottom}: The flux-ratio spectra of low- to high-mass mean spectra for the early-time (in red) and near-peak (in green) samples, respectively. The solid circles represent the mean flux ratios in bins of wavelength. As in Figure~\ref{mean-spec}, all of the spectra here are rebinned with a bin size of 15\,\AA.}
        \label{mean-spec-premax}
\end{figure}

In this work, we have shown that the near-peak UV spectra are promising indicators of progenitor metallicity. However, theoretical models also predict that the progenitor metallicity should have a more significant effect at earlier epochs \citep{2000ApJ...530..966L}: the SN~Ia UV spectra are expected to show larger differences with progenitor metallicities, and diverge at longer wavelengths. This implies that the early-time UV spectra could be better metallicity indicators.

To test this, we build an early-time sample with the UV spectra observed $\sim1$\,week before peak brightness. Our early-time sample contains 14 SNe, with phases ranging from $-9$ to $-6$\,days (a mean phase of $-7.6$\,days). Following the procedure described in Section~\ref{sec:mean-spectra}, we produce the mean UV spectra based on $\deltam(B)$ and host-galaxy stellar mass. Here the stellar mass is used instead of metallicity as it not only greatly increases the number of our early-time sample (13 out of 14 SNe have their host mass measured, but only half of them have host metallicity available), but also is a good proxy for metallicity \citep[e.g.,][]{2004ApJ...613..898T}. We then compare the flux-ratio spectra between the early-time and near-peak samples. The results are shown in Figure~\ref{mean-spec-premax}.

We find no significant difference between the slower decline and faster decline samples (here defined as $\deltam(B)<1.05$ and $\deltam(B)>1.05$\,mag, respectively), indicating that the decline rate is likely to have a negligible effect on the early-time UV spectra. However, there is an obvious trend that the mean spectra between low-mass and high-mass host galaxies present larger differences at $t \approx -7$\,days than that at peak brightness (see also Figure~\ref{mean-spec}). The spectra also tend to diverge at longer wavelengths. Using the stellar mass as the proxy for metallicity, this implies that the early-time UV spectra are better indicators of progenitor metallicity. Our finding is consistent with the theoretical predictions, although with large uncertainties owing to the small number of SNe with early-time observations available. 

The fast turnaround of {\it Swift} makes it an ideal instrument for increasing the number of SNe with early-time UV spectra in the future. With earlier observations, we should be able to detect the metallicity differences more easily. This especially has the advantage of mitigating the problem that the SNR of {\it Swift} UV spectra generally deteriorates dramatically at shorter wavelengths, barely probing the region where metallicity just starts to take effect for the near-peak spectra (i.e., $\lambda \approx 2700$\,\AA).

\section{Conclusions}
\label{sec:conclusion}
Theoretical studies predicted that the SN~Ia UV spectrum is sensitive to the progenitor metallicity. However, it is extremely difficult to constrain such metallicity effects from observations owing to either limited wavelength coverage or sample size. In this work, we present the analysis of the largest UV ($\lambda<2900$\,\AA) spectroscopic sample of SNe~Ia observed by {\it Swift}. We constructed mean spectra and spectral templates to investigate the relation between UV spectra and both SN and host-galaxy properties.

\begin{enumerate}
\item[$\bullet$] We confirm previous findings that the UV flux shows a strong dependence on SN light-curve shape, with slower-decline SNe having higher flux levels at $\lambda\sim3000$\,\AA. There is also some potential (but opposite) trend with SN light-curve shape at $\lambda\lesssim2700$\,\AA.

\item[$\bullet$] Using host galaxy as a proxy for the SN progenitor system, we find a $\sim2\sigma$ trend between the UV spectrum and host metallicity. SNe found in more metal-rich galaxies (thus likely to have more metal-rich progenitors) tend to exhibit a lower UV flux level at $\lesssim2700$\,\AA. We find that this metallicity effect is nearly negligible at longer wavelengths. Our results are consistent with the theoretical predictions and provide strong observational evidence for a correlation between SN~Ia UV spectrum and progenitor metallicity.

\item[$\bullet$] Following \citet{2016MNRAS.461.1308F}, we generate UV spectral templates that depend on light-curve shape and host metallicity. While the UV templates with only a single parameterisation using light-curve shape agree quite well with the observations, they fail to describe the behaviour of the spectrum at shorter wavelengths (e.g., $\lambda\lesssim2700$\,\AA). The two-parameter templates which take both light-curve shape and metallicity into consideration are more consistent with the observations. Although other parameters could also play a role, the strong dependence of SN~Ia UV spectra on light-curve shape and metallicity implies that they are important factors.

\item[$\bullet$] We investigate the relation between UV spectra and SN luminosity as parameterised by Hubble residuals. We do not see a significant trend between UV spectra and Hubble residuals; if anything, the flux level at $\sim2500$\,\AA\ (which is sensitive to the progenitor metallicity) tends to increase with Hubble residuals. However, the determination of Hubble residuals for our sample suffers large uncertainties from SN distances, as most of our SNe are extremely nearby ($z\lesssim0.01$). That means they only weakly depend on the cosmological model. Future analyses of higher-$z$ SN samples will be necessary to examine the metallicity bias on Hubble residuals.

\item[$\bullet$] We find that early-time UV spectra are likely better metallicity indicators than near-peak UV spectra. We generate the mean spectra for SNe in low-mass and high-mass galaxies at earlier epochs ($t \approx -7$\,days) and find that their difference tends to be larger than that of the near-peak sample. This implies that the early-time UV observations could be better probes of progenitor metallicity.
\end{enumerate} 

UV spectra are critical to understanding the explosion mechanism and progenitor systems of SNe~Ia. In particular, they are expected to be more direct probes of SN progenitor metallicity, which is believed to be able to significantly alter the SN luminosity and bias the distance measurement. Future studies with rest-frame UV spectra for higher-$z$ SNe will be critical for investigating the metallicity evolution and potential bias on cosmological analyses.

\section{Acknowledgements}
\label{sec:acknowledgements}
{\it Swift} spectroscopic observations were performed under programs
GI--04047, GI--5080130, GI--6090689, GI--8110089, GI--1013136, and
GI--1215205; we are very grateful to N.\ Gehrels, S. B.\ Cenko, and the
{\it Swift} team for executing the observations quickly.
Based in part on observations made with the NASA/ESA {\it Hubble Space
  Telescope}, obtained at the Space Telescope Science Institute
(STScI), which is operated by the Association of Universities for
Research in Astronomy, Inc., under National Aeronautics and Space
Administration (NASA) contract NAS 5--26555. These observations are
associated with Programmes GO--12298, GO--12592, GO--13286, GO--13646,
and DD--14925.  This manuscript is based upon work supported by NASA
under Contract No.\ NNG16PJ34C issued through the {\it WFIRST} Science
Investigation Teams Programme.

Y.-C.P. is supported by the East Asian Core Observatories Association (EACOA)
Fellowship. N.P.M.K. has been supported by the UK Space Agency.  
The UCSC group is supported in part by NASA grant 14-WPS14-0048, NSF grant
AST-1518052, the Gordon \& Betty Moore Foundation, and by fellowships
from the Alfred P.\ Sloan Foundation and the David and Lucile Packard
Foundation to R.J.F. Additional financial assistance to A.V.F. was
provided by the TABASGO Foundation, the Christopher R. Redlich Fund,
and the Miller Institute for Basic Research in Science (U.C. Berkeley).

The national facility capability for SkyMapper has been funded through ARC LIEF grant LE130100104 from the Australian Research Council, awarded to the University of Sydney, the Australian National University, Swinburne University of Technology, the University of Queensland, the University of Western Australia, the University of Melbourne, Curtin University of Technology, Monash University, and the Australian Astronomical Observatory. SkyMapper is owned and operated by The Australian National University's Research School of Astronomy and Astrophysics. The survey data were processed and provided by the SkyMapper Team at ANU. The SkyMapper node of the All-Sky Virtual Observatory (ASVO) is hosted at the National Computational Infrastructure (NCI). Development and support the SkyMapper node of the ASVO has been funded in part by Astronomy Australia Limited (AAL) and the Australian Government through the Commonwealth's Education Investment Fund (EIF) and National Collaborative Research Infrastructure Strategy (NCRIS), particularly the National eResearch Collaboration Tools and Resources (NeCTAR) and the Australian National Data Service Projects (ANDS).

\bibliographystyle{mn2e}
\bibliography{swift_ia_analysis}

\label{lastpage}

\end{document}